%% file: main.tex
\documentclass{article} %

\usepackage{hyperref}
\usepackage{url}
\usepackage{booktabs}       %
\usepackage{amsfonts}       %
\usepackage{nicefrac}       %
\usepackage{microtype}      %
\usepackage{amsmath,amsfonts,amsthm}       %
\usepackage{subcaption}
\usepackage{bm}
\usepackage{bbm}
\usepackage{multirow}
\usepackage{multicol}
\usepackage[inline]{enumitem}
\usepackage{diagbox}
\usepackage[capitalise]{cleveref}  %
\usepackage{comment}
\usepackage{etoolbox}
\usepackage{graphicx}
\usepackage{wrapfig}
\usepackage[font=small]{caption}
\usepackage{subcaption}
\usepackage{tablefootnote}
\usepackage{capt-of}
\usepackage{varwidth}

\usepackage{pifont}%

\newtheorem{exmp}{Example}[subsection]
\newtheorem{remark}{\bf Remark}[section]
\newtheorem{definition}{Definition}[section]
\newtheorem{proposition}{Proposition}[section]
\newtheorem{corollary}{Corollary}[section]

\usepackage[square,numbers,sort]{natbib}
\newlength{\defbaselineskip}
\usepackage[dvipsnames]{xcolor}         %

\usepackage[ruled,vlined]{algorithm2e}

\usepackage{tikz}
\usetikzlibrary{positioning,angles,quotes,matrix,shapes}

\DeclareMathOperator*{\argmin}{\arg\!\min}
\newcommand{\busset}{\mathbf{V}}
\newcommand{\rbusset}{\mathcal{E}}
\newcommand{\busedges}{\mathbf{E}}

\newsavebox\verbbox

\title{Field Teams Coordination for Earthquake-Damaged Distribution System Energization\footnote{
  This is an accepted manuscript version of an article published in the journal \href{https://www.sciencedirect.com/journal/reliability-engineering-and-system-safety}{``Reliability Engineering \& System Safety"}, doi: \url{https://doi.org/10.1016/j.ress.2024.110050}.
}}
\usepackage{authblk}
\author[]{\.{I}lker I\c{s}{\i}k}
\author[]{Ebru Aydin Gol}
\affil[]{Department of Computer Engineering, Middle East Technical University}
\affil[]{{\texttt{iilker@metu.edu.tr}, \texttt{ebrugol@metu.edu.tr}}}
\date{}

\begin{document}
\maketitle

\begin{abstract}
  The re-energization of electrical distribution systems in a post-disaster scenario is of grave importance as most modern infrastructure systems rely heavily on the presence of electricity.
This paper introduces a method to coordinate the field teams for the optimal energization of an electrical distribution system after an earthquake-induced blackout.
The proposed method utilizes a Markov Decision Process (MDP) to create an optimal energization strategy, which aims to minimize the expected time to energize each distribution system component.
The travel duration of each team and the possible outcomes of the energization attempts are considered in the state transitions.
The failure probabilities of the system components are computed using the fragility curves of structures and the Peak Ground Acceleration (PGA) values which are encoded to the MDP model via transition probabilities.
Furthermore, the proposed solution offers several methods to determine the non-optimal actions during the construction of the MDP and eliminate them in order to improve the run-time performance without sacrificing the optimality of the solution.\footnote{The code is publicly available at \url{https://github.com/necrashter/PowerRAFT}.}
\end{abstract}

\input{sections/intro}

\input{sections/prelim}

\input{sections/mdp}

\input{sections/experiment}

\input{sections/conclusion}

\bibliography{biblio}

\newpage

\appendix

\input{sections/proposition-proof}
\input{sections/corollary-proof}

\end{document}

%% file: sections/intro.tex
\section{Introduction}


Natural hazards that can result in disasters are part of life and can not be prevented. Prior planning and preparing management strategies are essential actions, among many others, to reduce the effects of a disaster.  Due to modern civilization's dependence on electricity, efficient power system restoration is crucial for disaster management~\citep{preweather}. 
Typically, an electrical distribution system experiences a complete blackout in the aftermath of an earthquake; all circuit breakers open in order to protect the infrastructure. The black-start of the distribution system is a challenging problem that requires the consideration of several electrical and topological constraints, even without any possible structural damage. An energization strategy for a black-start problem defines actions to re-connect each component to an energy source. In order to energize a component, there has to be an energized path (possibly through other components) to an energy source.  
In a simplified view, the components are energized step by step, starting from the components that have a direct connection to an energy source.
The decision-maker needs to orchestrate this black-start process in such a way that the time to energize each component is minimized while coping with the possibility that some field instruments may be damaged and non-operational due to the earthquake. In this paper, we present an MDP-based approach to generate optimal routes for the field teams to re-energize the distribution system after an earthquake-induced blackout.


The power system restoration has been studied from various aspects~\citep{prehurricane,preweather,back6,back7,back8,back9,Nunavath2019,Tan2020,Song2020,Hajian2016, Xie}. 
The optimal restoration plans based on electrical constraints using the sensory data are presented in~\cite{preweather, back6, back7, back8, back9}. The time required to repair damaged components is also integrated into the optimization in~\cite{prehurricane}.
Recently, decision-making approaches using artificial intelligence have also been applied in disaster management~\cite{Nunavath2019,Tan2020,Song2020,Hajian2016, Xie}. 
None of the aforementioned works consider the probabilistic health information of the field components or the limitations and constraints imposed by the transportation of the field teams that need to visit the field component to perform the energization action. In particular, it is assumed that the health status of each component is known either via field observation or sensory feedback. 
However, it would take a long time to obtain the health status of the whole network, and the sensor infrastructure is not commonly in place.

The coordination of the field teams after a disaster for restoration and repair purposes is studied in~\cite{Li2021,Ding2020,Yan2020,Arif}. A multi-stage algorithm starting with an initial route for the field teams in the first stage, and then updating the route in a stochastic manner as more information is received in the following stages is given in~\cite{Li2021}. Coordination of repair teams and the mobile battery-carrying vehicles via solving a Mixed Integer Linear Programming (MILP) problem is studied in~\cite{Ding2020}. 
The change in the repair duration with respect to the time is also modeled in~\cite{Yan2020}. The two-stage algorithm introduced in~\cite{Arif} deploys the field teams to the damaged components in the first stage and solves the optimal restoration problem while considering the uncertainties in the repair times and forecasts in the second stage. 
All the aforementioned works~\cite{Li2021,Ding2020,Yan2020,Arif}, address restoration and repair simultaneously, assuming that the list of non-operational components is available, which may require a significant amount of time to obtain following a disaster.
 
The study by Sharma~\cite{Sharma} presents a multistage framework for resilience analysis and recovery optimization. The framework considers the probabilistic health information and various factors affecting recovery, such as repair durations and the capabilities of the field teams. Similar to our work and resilience analysis presented in~\cite{Iannacone,Tian}, the framework from~\cite{Sharma} relies on the peak ground acceleration (PGA) values recorded during the earthquake \citep{seismic} for assessing the probabilistic health statuses of the components. However, in contrast to their focus on repair scheduling~\cite{Sharma} and resilience analysis~\cite{Sharma,Iannacone,Tian}, our approach involves determining the optimal routes for field teams to carry out energization actions exclusively. 
Determining optimal routes for restoration and recovery is studied in the context of fault localization. In~\cite{Fanucchi} multi-objective optimization algorithms are employed to generate the optimal route for the field team. In~\cite{Fanucchi2019}, the historical data is utilized to generate the route following the remotely performed actions for fault localization. 
In these prior studies, energization actions are remotely executed, and the route is generated for a single team. In contrast, our work addresses scenarios where, either due to inadequate infrastructure or earthquake-induced effects, the remote actions are not feasible, and multiple teams are coordinated to energize and reconfigure the system when the faults are known probabilistically.

The restoration policy generation presented in \cite{Gol2019, Arpali2019,ifacPriority,Yilmaz2023} utilizes the probabilistic health information by using Markov Decision Processes (MDPs), which provide a mathematical framework for modeling decision-making in situations where the outcome is affected by randomness.
Specifically, \cite{Gol2019} uses the PGA values recorded during the earthquake to compute the failure probabilities of the field components.
Then, these probabilities are used to construct an MDP model for the restoration problem. Furthermore, electrical and topological constraints are embedded in the MDP model, guaranteeing that the resulting sequence of restoration actions is applicable. Given that the costs in the MDP model are correctly adjusted, synthesizing an optimal policy for the MDP yields an optimal restoration strategy for the distribution system.
\cite{Arpali2019} extends \cite{Gol2019} by defining the state cost as the number of unenergized components.
This formulation minimizes the expected restoration time for the system. 
\cite{ifacPriority} further extends \cite{Gol2019} by prioritizing the components during the restoration, e.g., such as the components feeding the hospitals or the critical infrastructure.
Lastly, \cite{Yilmaz2023} presents an efficient online strategy update algorithm for the MDP formulation from \cite{Gol2019} in order to incorporate the information received from the field and the results of power flow analysis.
 However, the limitations and the time required to transfer the field teams are not considered in~ \cite{Gol2019,Arpali2019,ifacPriority,Yilmaz2023}. 
Instead, each restoration action is assumed to take a unit amount of time, i.e., either a remote control system is available and operational after the disaster, which is not common in electrical distribution networks, or the travel time between any two components is the same, which is quite unrealistic.
In the absence of a remote control system, the methods proposed in these papers fail to provide an optimal restoration strategy. In this paper, we build on the approach presented in \cite{Arpali2019} and integrate the field teams' mobility into the MDP model.  To the best of our knowledge, no other work solves the distribution system restoration problem by considering the mobility of the field teams, the electrical and topological constraints of the distribution system and the probabilistic health information of the components.

In this work, we extend the state space and the action space of the MDP from \cite{Arpali2019} to integrate the locations of the field teams, their travel durations, and the routes assigned to them. 
By synthesizing a policy for the new MDP model, we solve the restoration problem with field teams optimally. 
The proposed extension significantly inflates the MDP model's state space compared to~\cite{Arpali2019}. To overcome the related computational issues, we define a reduced MDP such that the optimal policy generated for the reduced model maps to the optimal policy of the original model. As the size of the reduced model is, in general, significantly less than the original model, the proposed reduction allows us to apply the approach in realistic scenarios. We further reduce the model size by eliminating the non-optimal actions during the MDP construction. As a result, the performance of the algorithm is improved significantly without sacrificing the optimality of the solution, which is illustrated over benchmarks.

The major contribution of this paper is an MDP-based approach for generating an optimal restoration strategy for a distribution network after an earthquake by considering the probabilistic health information of the components using the real-time earthquake data and the mobility of the field teams. The proposed methods are implemented in a tool called \textbf{PowerRAFT: Power Restoration Application with Field Teams}. The tool and the benchmarks are available at \url{https://github.com/necrashter/PowerRAFT}.

\textit{Organization:} The preliminary information on MDPs and distribution networks is given in Section~\ref{sec:prelim}. The proposed MDP model and the performance optimization techniques are given in Sections~\ref{sec:model} and~\ref{sec:opt}. The experimental results are presented in Section~\ref{sec:experiment}. Finally, the paper is concluded in Section~\ref{sec:conclusion}.

%% file: sections/prelim.tex

\section{Preliminaries and Notation}
\label{sec:prelim}

\subsection{Markov Decision Process}\label{sub:MDP}


\begin{definition}
A Markov Decision Process is a tuple $M = (S, A, p, c)$ with the following components~\citep{BertDynamic}:
\begin{itemize} \setlength{\parskip}{0pt}
  \setlength{\itemsep}{1pt}
    \item $S$ is a finite set of states.
    \item $A$ is a finite set of actions.
    \item $p(s' \mid s, a)$ is the probability of reaching  $s' \in S$ when action $a \in A$ is applied in state $s \in S$.
    \item $c: S \times A \times S \rightarrow \mathbb{R}^{0+}$ is the cost function that maps each combination of predecessor state, taken action, and successor state to a cost value.\end{itemize}
\end{definition}

A deterministic MDP policy $\pi: S \to A$ determines the action to be taken in each state. 
For a given policy $\pi$, the n-step value function $V_n^{\pi}(s)$ represents the expected value of the accumulated cost when the decision-maker complies with the policy $\pi$ for n-steps. The recursive definition of $V_n^{\pi}(s)$ for $n>0$ is as follows:
\begin{equation}\label{def:value}
  V_n^{\pi}(s) = \sum_{s' \in S} p(s' \mid s, \pi(s)) ( c(s, \pi(s), s') + V_{n-1}^{\pi}(s') )
\end{equation}
with the base condition $V_0^{\pi}(s) = 0$. Typically, the goal of a policy synthesis problem is to find the optimal policy $\pi^*$ that minimizes the value ($V_n^{*}(s)$ for each $s \in S$).
The value function of this policy can be computed using \eqref{def:opvalue} for $n>0$.
\begin{equation}\label{def:opvalue}
        V_n^{*}(s) = \min_{a \in A} \sum_{s' \in S} p(s' \mid s, a) ( c(s, a, s') + V_{n-1}^{*}(s') )
\end{equation}
with the base condition $V_0^{*}(s) = 0$. 
It is straightforward to extract the optimal policy $\pi^*(s)$ from $V_n^{*}(s)$ as shown in~\eqref{def:oppol} (i.e., the action $a$ minimizing $V_n^{*}(s)$ for each state and time step pair). 
\begin{equation}\label{def:oppol}
        \pi^{*}(s) = \argmin_{a \in A} \sum_{s' \in S} p(s' \mid s, a) ( c(s, a, s') + V_{n-1}^{*}(s') )
\end{equation}

Finally, a state $s_t \in {S}$ is called terminal if the only applicable action leads to the same state with a probability of $1$, i.e., $p(s_T \mid s_T, a) = 1$ for an action $a \in A$ and no other action is applicable in $s_T$.
Intuitively, once the MDP enters a terminal state, it cannot leave it.

\subsection{Restoration Process}\label{sub:restoration_prelim}


  \subsubsection{Probability of Failure ($P_f$) Values}\label{sub:Pf}


  A fragility curve is a function that maps the earthquake intensity level to the probability of exceeding a damage level~\citep{seismic}.
  Fragility curves have been studied extensively in the literature to estimate a structure's resilience against seismic damage~\citep{seismic,CIMELLARO20103639,multihazard_risk_assesment}.
  These curves are obtained by analyzing the response of a given structure to various levels of seismic excitation in a simulated environment.
  The earthquake intensity level is represented by the peak ground acceleration (PGA) value, which is calculated from the recorded earthquake data. In this work, the fragility curves are used to determine the probability of failure values ($P_f$) of distribution system components.

\subsubsection{Distribution System}\label{sub:distribution_system}

A distribution system with $N$ buses and a single energy source with infinite capacity (transmission grid) is represented as a tuple $\mathcal{DS} = (\busset, \rbusset, \busedges)$ where $\busset=\{1,\ldots,N\}$ is the set of buses, $\rbusset \subseteq \busset$ is the set of buses that are directly connected to the energy source, $\busedges \subseteq \busset \times \busset$ is the set of branches (physical connections) between the buses.
The pair of $\busset$ and $\busedges$ constitute an undirected graph $G = (\busset, \busedges)$.
In particular, if $(i,j) \in \busedges$, then $(j,i) \in \busedges$.

\begin{remark}
	Alongside the transmission grid, a distribution system can be energized from various distributed energy resources (DERs), such as generators and renewable energy resources. 
	Such resources are omitted to keep the notation simple, but they can be easily incorporated into the proposed solution, as in~\citep{Arpali2019,Gol2019}.
\end{remark}

We assume that all circuit breakers (switches) are open after an earthquake.
The health status of each bus is unknown at the start. For each bus $i$, $P_f(i)$ represents the probability that bus-$i$ will turn out to be damaged.
To re-energize the area, the breakers are closed by the field teams or remotely from the control center (if such a system is in place).
The actual health status is inferred from the result of the energization action.
Thus, the status of a bus $i$ is either \textit{unknown}, \textit{damaged}, or \textit{energized} for restoration purposes.

\begin{exmp}
	\label{ex:distribution_system}

	\begin{figure}[htbp]
		\centering
		\begin{minipage}{.49\textwidth}
			\centering
			\includegraphics[width=.8\textwidth]{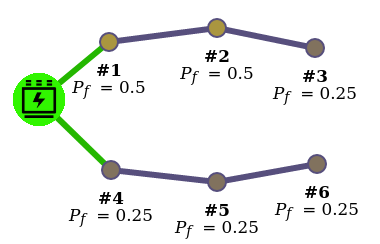}
			\caption{Sample distribution system}
			\label{fig:distribution_system_example}
		\end{minipage}
		\begin{minipage}{.49\textwidth}
			\centering
			\includegraphics[width=.8\textwidth]{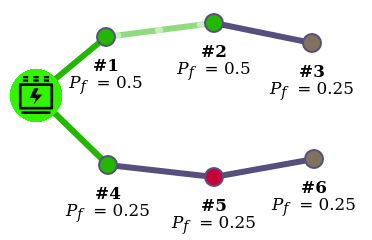}
			\caption{Sample configuration}
			\label{fig:distribution_system_state}
		\end{minipage}
	\end{figure}

	\begin{figure}[htbp]
	\end{figure}

	A sample distribution system is displayed in Figure \ref{fig:distribution_system_example}.
	The large green node represents a substation (through which the buses are connected to the transmission grid), whereas other nodes represent the buses. 
	Underneath each bus, its index and failure probability ($P_f$) are given.
	Buses 1 and 4 are connected to the substation.
	The edges in the graph represent the branches between buses and energy sources.

	The distribution system is represented by the tuple $\mathcal{DS} = (\busset, \rbusset, \busedges)$ with the following components:
	\begin{gather}
		\busset = \{1,2,3,4,5,6\} , \quad \rbusset = \{ 1, 4 \} \\
		\busedges = \{
			(1,2), (2,1),
			(2,3), (3,2),
			(4,5), (5,4),
			(5,6), (6,5)
		\}
	\end{gather}

  The probability of failure ($P_f$) for each bus $i \in \{1,2,3,4,5,6\}$ is defined as follows:
  
	\begin{align}
  	P_f(i) = \begin{cases}
			0.5    & i = 1 \text{ or } i = 2 \\
			0.25   & \text{otherwise}
		\end{cases}
	\end{align}

	As shown in Figure~\ref{fig:distribution_system_example}, all buses are in unknown status after an earthquake. The color of such a bus is interpolated between gray and yellow depending on its failure probability.
  Another sample distribution system configuration is given in Figure~\ref{fig:distribution_system_state}.
	Energized buses are shown with green, and damaged buses are shown with red.
\end{exmp}

\subsubsection{Field Teams Mobility}
\label{sub:FTM}

A bus that is not remotely controlled can only be energized by an on-site field team. A restoration strategy determines the target buses for field teams, considering the most recent information about the system and team locations throughout the restoration process. 

We assume that the time required to travel from a certain point to another is constant, discrete, and the same for all teams (i.e., all field teams travel at the same speed).
We also assume that the time required for an energization attempt is negligible (instantaneous) compared to the travel durations since it only involves closing a circuit breaker.

To represent the travel durations,
we assume we are given a function $time: \busset \times \busset \rightarrow \mathbb{N}$ that maps the pairs of buses to the time required to travel from the first one to the second.
We further assume that the $time$ function conforms to the following rules:
the triangle inequality \eqref{triangle_inequality} must hold,
and the function must yield a positive value \eqref{dist_to_another} unless both inputs are the same, in which case it must yield zero \eqref{dist_to_itself}.
\begin{align}
	& time(i,j) + time(j,k) \geq time(i,k), \forall i,j,k \in \busset \label{triangle_inequality} \\
	& time(i,j) > 0, \forall i,j \in \busset, i \neq j \label{dist_to_another} \\
	& time(i,i) = 0, \forall i \in \busset \label{dist_to_itself}
\end{align}
The $time$ function can account for other constraints, such as obstacles in the way and bad terrain, provided that these assumptions are not violated.
Also note that the codomain of the function $time$ is $\mathbb{N}$ since we assume that travel times are discretized. Any function that represents the continuous values in $\mathbb{R}^{0+}$ with approximate discrete integers can be used for discretization.
For example, we can simply round up to the nearest integer using $round(x) = \lceil x \rceil$.
Or we can discretize more coarsely, e.g., the function $round(x) = \lceil x/C \rceil$ where $C$ is a constant integer.
This results in a simpler, less precise model, which is faster to generate in practice.

\subsection{Problem Formulation}\label{sub:problem}

A restoration strategy defines the overall course of action to energize the distribution system. A strategy is formalized as a sequence of actions, or those with alternate plans, or as a network topology with the corresponding set of closed switches \cite{Qiu2017,back6}.
In this work, we define a \textit{restoration strategy} as a mapping from a distribution system configuration to an action to take in this state. 
Each action guides each team to a target bus to energize.

The optimal course of action to take depends on the goal of the restoration process, e.g., minimizing the average energization time, minimizing the maximum energization time, or prioritization of some components \cite{Arpali2019, Gol2019, ifacPriority}.
In this paper, we aim to minimize the expected time to energize each bus by considering the travel durations of field teams.

In this paper, our goal is to restore electricity to an earthquake-damaged distribution system $\mathcal{DS} = (\busset, \rbusset, \busedges)$ (Section~\ref{sub:distribution_system}) optimally.
We assume that $F$ teams are at the field to perform restoration actions on-site; the travel durations of the teams are represented by the $time$ function (Section~\ref{sub:FTM}), and the probability of failure $P_f$ values are computed for each system component after the realized earthquake (Section~\ref{sub:distribution_system}). Given these, our goal is to design a restoration strategy that minimizes the total expected restoration time.

To generate an optimal restoration strategy that considers the failure probabilities and the travel durations, we model the restoration operation as an MDP $M = (S, A, p, c)$. The configuration of the distribution system (energized, damaged, unknown) and the locations of the field teams are encoded in the states of the proposed MDP model ($S$). The $P_f$ values are integrated to the state transition probability function ($p$). An MDP action ($a \in A$) determines bus assignments to field teams. The cost ($c(s,a,s')$) is determined with respect to the number of unenergized buses.
Consequently,  minimizing the total expected cost reduces to minimizing the actual expected average restoration time for the distribution system. The model and the restoration strategy are explained in detail in the following section.

%% file: sections/mdp.tex
\section{Proposed MDP Model}
\label{sec:model}

In this section, the proposed MDP model $M = (S, A, p, c)$ for coordinating the field teams to restore the earthquake-damaged distribution system is explained in detail.

\subsection{States}
\label{sub:state}

The states $S$ of the proposed MDP model $M$ represent the configuration of the distribution system and the locations (en route / at a bus) of each field team. In particular,  each state $s = (\mathbf{s}, \mathbf{t}) \in S$ is a tuple, where $\mathbf{s} = [s_1, s_2, \ldots, s_N] $ is the distribution system configuration described in Section~\ref{sec:distribution_system_state}, and $\mathbf{t} = [t_1, t_2, \ldots, t_F]$ is the locations of the field teams described in Section~\ref{sec:field_teams_state}. 

\subsubsection{Distribution System State}\label{sec:distribution_system_state}

As described in Section~\ref{sub:distribution_system}, a bus can be in an \textit{unknown} (not observed or tried to be energized), \textit{damaged}, or \textit{energized} status. These are denoted by $U$, $D$, and $E$, respectively. The status of bus-$i$ is shown with $s_i \in \{U,D,E\}$, and the status of the whole system is shown with a vector $\mathbf{s} = [s_1, s_2, \ldots, s_N]$. Finally, the set of all possible states for the distribution system is defined as $\mathcal{S}$~\eqref{eqn:distribution_system_states}.
\begin{equation}
    \mathcal{S} = \{ \textbf{s} = [s_1, s_2, \ldots, s_N] \mid  s_i \in \{ D, U, E \}  \text{ for each } i \in \busset \} \label{eqn:distribution_system_states}
\end{equation}

\subsubsection{Field Teams State}\label{sec:field_teams_state}

The status of team-$i$ is denoted by $t_i$  in $\mathbf{t} = [t_1, t_2, \ldots, t_F]$. 
Each $t_i$ is a tuple $t_i = (t_{i,1}, t_{i,2})$, where $t_{i,1} \in \busset$ is the target bus and $t_{i,2} \in \mathbb{N}$ is the remaining time until the team reaches $t_{i,1}$.
A team can either be at a bus, in which case $t_{i,2} = 0$, or traveling from one bus to another one (en route), in which case $t_{i,2} > 0$.
The set of all possible states for a field team is defined as $\mathcal{T}$:

\begin{equation}
\label{def:T}
    \mathcal{T} = \busset \times \mathbb{N}
\end{equation}

Consequently,  the set of all possible team states is defined as $\mathcal{T}^F$ ( $\mathcal{T}^F = \mathcal{T}_1 \times  \ldots \times \mathcal{T}_F$, where $\mathcal{T}_i = \mathcal{T}$), and the set of MDP states $S$ is
\begin{equation}
    S = \mathcal{S} \times \mathcal{T}^F
\end{equation}
Lastly, the initial state of the MDP is given in~\eqref{eq:initial_state}, where the $t_{i,1} \in \busset$ is the initial location of the team-i and $s_{i,0}$ is $U$ for each bus $i \in \busset$. 
\begin{equation}\label{eq:initial_state}
  s_0 = ([s_{1,0}, \ldots, s_{N,0}], [(t_{1,1}, 0), \ldots, (t_{F,1}, 0)]), 
\end{equation}

\begin{remark}
\label{remark:startpos}
Presumably, not every team will be located near a bus at the beginning of the restoration process.
Such initial conditions can be represented by extending $\busset$ with new indices (e.g., buses that are not connected to any other buses or energy sources via branches). 
However, since such a trivial extension does not pose any technical challenges, we did not incorporate it to avoid further complicating the notation. 
\end{remark}

\begin{figure}[htbp]
	\centering
	\begin{minipage}{.49\textwidth}
		\centering
		\includegraphics[width=.8\textwidth]{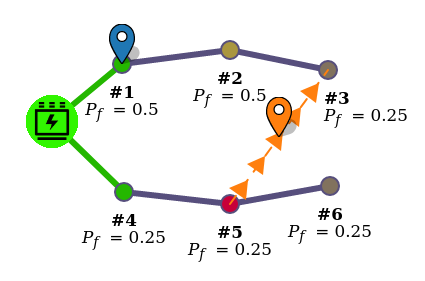}
    \caption{MDP state $\left(\textbf{s} , [(1, 0), (3, 1)] \right)$}
		\label{fig:state_example_1}
	\end{minipage}
	\begin{minipage}{.49\textwidth}
		\centering
		\includegraphics[width=.8\textwidth]{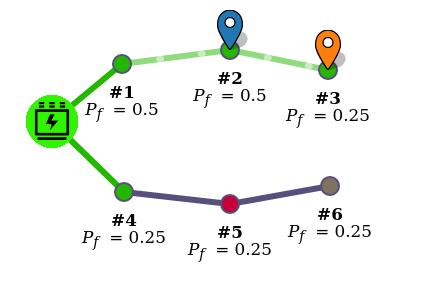}
    \caption{MDP state $\left(\mathbf{s'} , [(2, 0), (3, 0)] \right)$}
		\label{fig:state_example_2}
	\end{minipage}
\end{figure}

\begin{exmp}
\label{ex:state}
Sample MDP states for the distribution system introduced in Example~\ref{ex:distribution_system}, and two field teams are shown in Figures \ref{fig:state_example_1} and \ref{fig:state_example_2}.
Blue and orange markers denote the location of the first and the second teams, respectively.
An en route team is shown on an arrow that connects the source and the target buses.

The travel duration from bus-$5$ to bus-$3$ is 2 ($time(5,3)=2$), which means that the second team will arrive in 1 unit of time, thus $\mathbf{t} = [(1, 0), (3, 1)]$. Based on the color codes defined in Example \ref{ex:distribution_system}, the state of the distribution system is $\textbf{s} = [E, U, U, E, D, U]$ in Figure~\ref{fig:state_example_1}. In Figure \ref{fig:state_example_2}, both teams are located at the buses, thus $\mathbf{t'} = [(2, 0), (3, 0)]$ and the state of the distribution system is $\mathbf{s'} = [E, E, E, E, D, U]$. As there is no energizable path from the source to the bus-$6$ (the only bus in unknown status), this state is terminal.
\end{exmp}

\subsection{Action Set}
\label{sub:action_set}

An action $\textbf{a} = [a_1, a_2, \ldots, a_F] \in A$ of the MDP model defines a command for each team. We assume that when a team is at a bus, it tries to energize it if the electrical and topological constraints are satisfied, which are determined by the control center and further discussed in the next subsection.
Consequently, we only consider commands regarding traveling:  \textit{go to a bus} (denoted with the bus id) or \textit{wait} ($W$). Thus, the action set $A$ is defined as:

\begin{equation}
\label{def:A}
    A = \{ [a_1, a_2, \ldots, a_F] \mid  a_i \in \busset \cup \{ W \} \text{ for each } i = 1,\ldots,F \}
\end{equation}

\subsubsection{Energizable Buses}

Recall from Section~\ref{sub:distribution_system}, that an energization action can be applied to bus-$i$ if it is in unknown status (i.e., $s_i = U$) and it is connected to an energized bus ($s_j = E$ and $(i,j) \in \busedges$ for some $j \in \busset$) or an energy resource ($i \in \rbusset$). For a distribution system state $\mathbf{s}$, the set of all buses satisfying this pre-condition is shown with $\beta_1(\textbf{s})$ ~\eqref{eqn:beta1}.
Moreover, $\beta_n(\textbf{s})$ for $n > 1$ is defined as the set of buses that can be energized after $\beta_{n-1}(\textbf{s})$ is energized~\eqref{eqn:betan}.
The union of all non-empty $\beta_n(\textbf{s})$ for $n \geq 1$ yields $\beta(\textbf{s})$, the set of buses with unknown status and a non-blocked path of buses (i.e., a path without any buses that is known to be damaged) to an energy source~\eqref{eqn:beta}.
\begin{gather}
  \label{eqn:beta1}
  \beta_1(\textbf{s}) = \{ i \in \busset \mid s_{i} = U, i \in \rbusset \textrm{ or } ( s_j = E \textrm{ and } (i,j) \in \busedges \textrm{ for some } j \in \busset ) \}
  \\
  \label{eqn:betan}
  \beta_n(\textbf{s}) = \{ i \in \busset \mid s_{i} = U, s_j = E \textrm{ and } (i,j) \in \busedges \textrm{ for some } j \in \beta_{n-1}(\textbf{s}) \} \text{ for } n>1
  \\
  \label{eqn:beta}
  \beta(\textbf{s}) = \bigcup_{1 \leq n \leq N} \beta_n(\textbf{s})
\end{gather}
Note that 
$\beta_n(\textbf{s})$ for $n > N$ is guaranteed to be empty since there are $N$ buses and the length of the longest possible non-cyclic path in the distribution system graph is $N$.  Thus, $\beta(\textbf{s})$ is the set of buses that can be energized after a sequence of energization actions, which is used to define the target buses for the field teams. 

\begin{remark}
During the restoration, additional conditions for energization can be considered. Examples include constraints due to the capacity of an energy resource or constraints derived from power flow analysis. Any such constraint can be integrated into $\beta_1(\cdot)$~\ref{eqn:beta1} \cite{Arpali2019,Gol2019,Yilmaz2023}.
\end{remark}

\subsubsection{Feasible Actions for the Field Teams}

We use $A(s) \subset A $ to denote the set of feasible actions at state $s$ and $A_i(s)\subset \busset \cup \{ W \}$ to denote the set of feasible commands for the team-$i$ at state $s$. 
We assume that an en route team must continue traveling to its destination (it's not possible to re-route the team before reaching the destination).
If the team is at a bus, it can either wait (i.e., when it is not possible to energize it right away) or start traveling to another energizable bus ($j \in \beta(\textbf{s})$):

\begin{equation}
\label{def:AiS}
    A_i((\textbf{s},\textbf{t}))  =
    \begin{cases}
      \{ t_{i,1} \} & \textrm{ if } t_{i,2} > 0 \\
      \beta(\textbf{s})  \cup \{ W \text{ if } t_{i,1}  \in \beta(\textbf{s}) \} & \textrm{ if }  t_{i,2} = 0
    \end{cases}
\end{equation}

It is not possible to make an energization attempt if there are no buses with unknown status or all such buses are unreachable, \textit{i.e.}, all paths from energy resources to the buses with unknown status are blocked by the damaged system components. Observe that for such \textit{terminal} states $\textbf{s}$, it holds that $\beta_1(\textbf{s}) = \emptyset$~\eqref{eqn:beta1}. 
Re-routing the teams is not meaningful when such a state is reached for restoration purposes. Thus, in the MDP model, we introduce  only the wait action $\textbf{W} = [W,\ldots,W]$ as feasible for a state $s = (\textbf{s}, \textbf{t})$ with terminal distribution system part ($\textbf{s}$) regardless of the locations of the teams ($\textbf{t}$). For states with a non-terminal distribution part, $A(s)$ is the set of actions such that each team receives a command from its feasible set, and at least one team is commanded to travel a bus that can be energized: 

\begin{equation}
\label{def:AS}
    A((\textbf{s}, \textbf{t})) =
    \begin{cases}
    \{\mathbf{W}\} & if \beta_1(\textbf{s}) = \emptyset \\
    \begin{aligned}
		\{ [a_1, \ldots, a_F] \mid  \forall i \left ( a_i \in A_i(\textbf{s}, \textbf{t}) \right ) \textrm{ and }  \\
                  \exists {i} \left( a_i \in \beta_1(\textbf{s}) \right) \}
    \end{aligned}
    & \textrm{otherwise}
    \end{cases}
\end{equation}

Note that it is possible to send a team to a bus that can not be energized immediately (i.e., bus-$i$ with $i \not \in \beta_1(\textbf{s})$).
Such actions may be preferable because another team may energize a neighbor bus in the subsequent steps, thereby enabling the previous team to energize.
Since this depends on another team energizing a neighbor bus, the feasible set definition requires at least one team to be sent a bus from $\beta_1(\textbf{s})$.
This can be interpreted as a \textit{progress} requirement. Without this requirement, the teams could travel between buses that are not in $\beta_1(s)$ indefinitely. This results in loops in the MDP, which are avoided by the progress requirement for computational efficiency without sacrificing optimality.

\begin{exmp}\label{ex:action}
The sets of feasible actions for the teams are $A_1((\mathbf{s}, \mathbf{t})) = \{2,3,6\}$ and $A_2((\mathbf{s}, \mathbf{t})) = \{C\}$ for the MDP state $(\mathbf{s}, \mathbf{t})$ depicted in Figure~\ref{fig:state_example_1}. Since the second team is en route, it can only continue traveling, whereas the first team can be sent to any bus that is in unknown status.  However, $A((\mathbf{s}, \mathbf{t})) = \{ [2, C] \}$, since this is the only action that satisfies the progress condition.

The state in Figure~\ref{fig:state_example_2} is a terminal state since bus-6 is unreachable due to the damage of bus-5. Therefore, the only feasible action is $[W, W]$.
\end{exmp}


\subsection{Transitions}
\label{sub:transitions}

In this section, the state transition function $p$ of the MDP model is defined.  In particular, the states that can be reached with a non-zero probability when action $\mathbf{a} \in A$ is applied at state $s = (\mathbf{s}, \mathbf{t})$ and the corresponding probabilities are defined.
The locations of the field teams after one unit of time $\mathbf{t}'$ is deterministically defined with respect to the taken action $\mathbf{a}$ and the team locations $\mathbf{t}$.
Furthermore, based on their locations at $\textbf{t}'$, teams attempt to energize the system components that satisfy the topological and electrical constraints.
Due to the stochasticity of the health statuses, this yields several potential outcomes for the distribution system.
The probabilities of these outcomes are determined using failure probabilities of the system components that are attempted to be energized.

In the following sections, we first define the team locations $\mathbf{t}'$ that are reached from $\mathbf{t}$ in one step, then the possible energization attempts that can be applied in $\mathbf{s}$ with the field teams located at $\mathbf{t}'$, and finally the probability $p(s' \mid s, \mathbf{a})$ for any $(s,\mathbf{a},s') \in S \times A \times S$.

\subsubsection{Field Teams Transitions} \label{sub:team_transitions}

Given the statuses of the field teams $\mathbf{t} = [t_1, \ldots, t_F]$ and an action $\mathbf{a} = [a_1, \ldots, a_F] \in A((\mathbf{s},\mathbf{t}))$, the time remaining for team-$i$ to complete its command is given by the function $rem(\mathbf{t},\mathbf{a},i)$~\eqref{def:remSai} for ${a}_i \neq W$.
This function is undefined for the waiting teams since the waiting action can be sustained indefinitely.
For other teams, the remaining time is the number of time steps until the team reaches the target bus.

\begin{equation}
\label{def:remSai}
    rem(\mathbf{t},\mathbf{a},i) =
    \begin{cases}
    time(t_{i,1}, a_i) & \textrm{ if } t_{i,2} = 0 \\
    t_{i,2} & \textrm{ if } t_{i,2} > 0
    \end{cases}
\end{equation}

When the action $\mathbf{a}$ is applied at $\mathbf{t} = [t_1, \ldots, t_F]$, the positions of the field teams after one unit of time is $\mathbf{t}' = [t_1', \ldots, t_F']$:

\begin{equation}
\label{eqn:Tiprime}
    t'_i =
    \begin{cases}
    t_i & \textrm{if } {a}_i = W \\
    (a_i, rem(\mathbf{t},\mathbf{a},i) - 1)   & \textrm{otherwise}  \\
    \end{cases}
\end{equation}

Essentially, if the team is ordered to wait, the team stays at the same bus.
If the remaining time of the team is $1$, the team arrives at its destination, which is $a_i$.
Otherwise, the team progresses by 1 unit of time and remains en route.

\subsubsection{Distribution System Transitions} \label{sub:system_transitions}

Next, we define the set of distribution system states that can be reached with non-zero probability when action $\mathbf{a}$ is applied at state $s = (\mathbf{s}, \mathbf{t})$.
After the teams travel or wait for one unit of time, they attempt to energize the buses with respect to electrical and topological constraints. Note that these constraints are integrated into $\beta_1(\cdot)$~\eqref{eqn:beta1}, and the energization actions are given from a control center as the field teams do not necessarily know the whole status of the network. The teams report the result back to the control center or the status of the component via observation.
Consequently, the energization attempts are performed according to the positions $\mathbf{t}'$~\eqref{eqn:Tiprime}. In addition, a successful energization attempt at a bus-$i$ can enable other teams to energize a bus-$j$ that is connected to bus-$i$. We assume that the energization attempts are coordinated by the control center and performed iteratively until no further energization attempt can be performed considering the resulting distribution system state and the positions of the teams $\mathbf{t}'$. In the following, we first introduce the necessary notation to define the set of distribution system states that can be reached after such iterative energization attempts (denoted as $\Phi(\textbf{s}, \textbf{t}'$)), then compute the corresponding transition probability values with respect to the probability of failure values.

We first define the set of buses $\alpha(\mathbf{s}, \mathbf{t})$ that can be attempted to be energized in the distribution system state $\mathbf{s}$ with respect to the team locations $\mathbf{t}$. Naturally, this is similar to $\beta_1(\mathbf{s})$~\eqref{eqn:beta1} but restricted by the locations of the field teams.
The buses in the set $\alpha(\mathbf{s}, \mathbf{t})$ are referred to as ``energizable" ($\mathbf{t} = [t_1, \ldots, t_F]$).

\begin{align}
\label{def:alpha}
    \alpha(\textbf{s}, \textbf{t}) & = \beta_1(\textbf{s}) \cap \{ t_{i,1} \mid 1 \leq i \leq F, t_{i,2} = 0 \}  
\end{align}

After an energization attempt, the bus can turn out to be damaged ($D$) or become energized ($E$) by a neighbor component or an energy resource it is connected to.
For a distribution system state $\mathbf{s} = [s_1, \ldots, s_N]$, the set of all possible distribution system states after all the buses in $\alpha(\mathbf{s}, \mathbf{t})$ are attempted to be energized is defined in~\eqref{def:stateEnergize}.
\begin{equation}
    \label{def:stateEnergize}
    \phi(\mathbf{s}, \mathbf{t}) = \{ \mathbf{s}' \mid \forall i
    \begin{cases}
		s_i' \in \{E, D\} & \text{ if } s_i \in \alpha(\mathbf{s}, \mathbf{t}) \\
        s_i' = s_i & \text { otherwise}
    \end{cases}
    \}
\end{equation}

The set of distribution system states $\Phi(\mathbf{s}, \mathbf{t})$ that can be reached during the iterative energization attempts when the field teams are located at $\mathbf{t}$ is computed iteratively via the converging sequence
\[ \Phi_{1}(\mathbf{s}, \mathbf{t}) \subseteq  \Phi_{2}(\mathbf{s}, \mathbf{t}) \subseteq \ldots \]
where $\Phi_{1}(\mathbf{s}, \mathbf{t}) = \phi(\mathbf{s}, \mathbf{t})$ and 
\begin{align}
    \Phi_{n}(\mathbf{s}, \mathbf{t}) &= \Phi_{n-1}(\mathbf{s}, \mathbf{t}) \cup \bigcup_{\mathbf{s}' \in \Phi_{n-1}(\mathbf{s}, \mathbf{t})} \phi(\mathbf{s}', \mathbf{t}) \text{ for } n > 1 
\end{align}
The sequence is indeed converging since there is a finite number of distribution system states (for some $m \geq 1$, $\Phi(\mathbf{s}, \mathbf{t}) = \Phi_{m}(\mathbf{s}, \mathbf{t})$; and for each $l > 0$, $\Phi_{m}(\mathbf{s}, \mathbf{t}) =  \Phi_{m+l}(\mathbf{s}, \mathbf{t})$). A distribution system state can be reached after the iterative process if it is not possible to perform additional energization attempts. Note that such a state can be reached at any step of the iterative process. The set of such states is defined as:
\begin{equation}
    \overline{\Phi(\mathbf{s}, \mathbf{t}) } = \{\mathbf{s} \in \Phi(\mathbf{s}, \mathbf{t})  \mid \alpha(\mathbf{s}, \mathbf{t}) = \emptyset\}
    \end{equation}

Finally, the probability of reaching $s' = (\mathbf{s}', \mathbf{t}')$ when action $\mathbf{a}$ is applied at $s = (\mathbf{s}, \mathbf{t})$ is given in~\eqref{eqn:p} for MDP states $s' = (\mathbf{s}', \mathbf{t}') \in S$ with $\mathbf{t'}$ computed as in~\eqref{eqn:Tiprime} and $\mathbf{s}' \in \overline{\Phi(\mathbf{s}, \mathbf{t}')}$. For all other states  $s' \in S$, $p(s' \mid s, \mathbf{a}) = 0$. Observe that $s_i$ is $U$ and $s_i' \in \{ D, E\}$ when $s_i \neq s_i'$   due to~\eqref{eqn:beta} and \eqref{def:alpha}.
\begin{equation}
\label{eqn:p}
    p(s' \mid s, \mathbf{a}) = \prod_{i, s_i \neq s_i'} \begin{cases}
    P_f(i) & \textrm{ if } s'_i = D \\
    1 - P_f(i) & \textrm{ if } s'_i = E
    \end{cases}
\end{equation}

\begin{exmp}

Consider the MDP state $(\mathbf{s}, \mathbf{t})$ depicted in Figure~\ref{fig:state_example_1} and let $\mathbf{a} = [2,C] \in A((\mathbf{s}, \mathbf{t}))$ (defined in Example~\ref{ex:action}) be given in $(\mathbf{s}, \mathbf{t})$ and $time(1,2)$ be $1$. The remaining travel time is 1 for both teams, i.e., $rem(\mathbf{t}, \mathbf{a}, i)=1$. As they will both reach their targets after taking this action, the resulting locations are $\mathbf{t'} = [2,3]$~\eqref{eqn:Tiprime}. The following sets are computed during the iterative process: (where $ \mathbf{s}_1 = [E, E, U, E, D, U]$, $\mathbf{s}_2 = [E, D, U, E, D, U]$, $\mathbf{s}_3 = [E, E, E, E, D, U]$, $\mathbf{s}_4 = [E, E, D, E, D, U]$) 
\begin{align*}
& \alpha(\mathbf{s}, \mathbf{t}') = \{2\}, & \phi(\mathbf{s}, \mathbf{t}') = \{ \mathbf{s}_1,  \mathbf{s}_2\},   &   \quad \Phi_1(\mathbf{s}, \mathbf{t}') = \{ \mathbf{s}_1,  \mathbf{s}_2\} \\
&  \alpha(\mathbf{s}_1, \mathbf{t}')  = \{3\} ,  \alpha(\mathbf{s}_2, \mathbf{t}')  = \{\}  &   \phi(\mathbf{s}_1, \mathbf{t}') = \{ \mathbf{s}_3,  \mathbf{s}_4\}  & \quad   \Phi_2(\mathbf{s}, \mathbf{t}') = \{\mathbf{s}_1, \mathbf{s}_2, \mathbf{s}_3 , \mathbf{s}_4 \} \\
& \alpha(\mathbf{s}_3, \mathbf{t}')  = \{\}, \alpha(\mathbf{s}_4, \mathbf{t}')  =    \{\},  &    \phi(\mathbf{s}_3, \mathbf{t}') = \phi(\mathbf{s}_4, \mathbf{t}') = \{\} & \quad   \Phi(\mathbf{s}, \mathbf{t}')  = \Phi_2(\mathbf{s}, \mathbf{t}') 
\end{align*}
The series converges after the second iteration and $\overline{ \Phi(\mathbf{s}, \mathbf{t}') } = \{\mathbf{s}_2, \mathbf{s}_3, \mathbf{s}_4\}$. The corresponding transition probabilities are computed with respect to the $P_f$ values shown in Figure~\ref{fig:state_example_1}, and they are $p((\mathbf{s}_2, \mathbf{t'}), \mathbf{a}, (\mathbf{s}, \mathbf{t})) = 0.375$, $p((\mathbf{s}_3, \mathbf{t'}), \mathbf{a}, (\mathbf{s}, \mathbf{t})) = 0.375 $ and $p((\mathbf{s}_4, \mathbf{t'}), \mathbf{a}, (\mathbf{s}, \mathbf{t})) = 0.25$.
	The MDP state reached when all energization attempts are successful (i.e. $(\mathbf{s}_3, \mathbf{t'})$) is shown in Figure~\ref{fig:state_example_2}.

\end{exmp}

\subsection{Cost Formulation}
\label{sub:cost}
The cost of transitioning from state $(\mathbf{s}, \mathbf{t})$ to $(\mathbf{s}', \mathbf{t}')$ by taking action $\mathbf{a} \in A(s)$ is  the number of unenergized buses in $\mathbf{s} = [s_1,\ldots, s_N]$ (see~\eqref{def:c}).
A similar cost formulation was used in \cite{Arpali2019}.
\begin{equation}
\label{def:c}
    c((\mathbf{s}, \mathbf{t}), \mathbf{a}, (\mathbf{s}', \mathbf{t}')) = |\{s_i \mid s_i \in \{U, D\}, i=1, \ldots N\}|
\end{equation}

The total energization duration (within the optimization horizon) for each bus is penalized when the cost~\eqref{def:c} is used in combination with the finite horizon value function~\eqref{def:value}. Consequently, the optimal restoration strategy (policy)~\eqref{def:opvalue} minimizes the expected time to energize each bus.

\subsection{Optimization Horizon}

Given a state $s$, let $l$ be the longest path from $s$ to a terminal state.
We must look forward at least $l$ steps to determine the optimal action in $s$, i.e., the action extracted from $V_l^{*}$ is optimal for $s$.
For $n > l$, $V_n^{*}(s) = V_l^{*}(s)$.
Therefore, the minimum optimization horizon is the length of the longest path from the initial state to a terminal state.
Note that due to the progress condition~\eqref{def:AS}, the MDP has no loops, and the length of the longest path is finite.


\section{Performance Optimization}\label{sec:opt}

The proposed MDP model for the synthesis of the optimal restoration strategy encodes all possible restoration scenarios, which naturally results in a large model. In particular, the number of the states of the MDP model is exponential in the number of buses and the number of field teams. The total number of states is upper bounded by $3^N \times (N \times \overline{time})^F$ where $N$ is the number of buses, $\overline{time} = \max_{a,b \in \busset} time(a,b) - 1$, and $F$ is the number of field teams.
However, this theoretical bound is not reached in practice since many of these states are infeasible (e.g., includes an energized bus that is not connected to a source or another energized bus). 
Such infeasible states are not added to the MDP model when the model is constructed iteratively via one-step reachability analysis, i.e., starting from the initial state $s_0$ ($S = [s_0]$) and iteratively adding states to $S$ that are one-step reachable from a state that is already in $S$. 
Furthermore, dynamic programming is used during both the MDP construction phase (to avoid exploring the same state multiple times) and the policy synthesis (in value iteration).
Nevertheless, the number of states might still be large due to the number of different restoration strategies. 

In this section, we define two optimization approaches to reduce the total number of states in the resulting model. The first one allows us to process multiple transitions simultaneously, during which no energization attempts occur by embedding the travel times within transitions and modifying the cost and value functions accordingly. 
The second one allows us to discard the feasible actions that are guaranteed to be non-optimal and the corresponding successor states. Such actions cannot be part of an optimal policy. Both approaches potentially create an MDP model $M'$ with fewer states than the original model $M$. However, the optimal policies, thus the restoration strategies, are guaranteed to be the same.  

\subsection{Eliminating Deterministic Transitions}
\label{sub:modval}
If an energization attempt does not occur after taking an action, the transition is deterministic, and the successor state is determined by the team transitions. By combining the consecutive deterministic transitions into a single transition, the total number of states is reduced. In particular, since at least one team has to arrive at a bus with unknown status for an energization attempt to occur, increasing the amount of time passed in a transition so that at least one team arrives at their destination allows us to eliminate the deterministic transitions and the corresponding intermediate states. Note that the only feasible action in such an intermediate state is to continue to travel for all teams. Thus, eliminating those states, updating the corresponding transitions, and embedding the time passed into the cost and value functions result in the same strategy without the intermediate states. Next, this optimization approach is explained through a new value function ($\mathbb{V}$)  and cost function $c_n$.

Given a state $s = (\mathbf{s}, \mathbf{t})$ and an action $\mathbf{a}$, the time required for a team in $s$ to arrive at its destination is defined in~\eqref{eqn:tSaW}. 
 \begin{align}
    t((\mathbf{s}, \mathbf{t}),\mathbf{a}) & = \min_{1 \leq i \leq F, a_i \neq W} rem(\mathbf{t},\mathbf{a},i) \quad\textrm{ when } \mathbf{a} \neq \mathbf{W}  \nonumber \\
    t((\mathbf{s}, \mathbf{t}),\mathbf{a}) & = 1 \quad \textrm{ when } \mathbf{a} = \mathbf{W}
\label{eqn:tSaW}
\end{align}

Fundamentally, we find the time required to complete the action for each non-waiting team if there is at least one such team, and the minimum of these values is the required amount of time for at least one team to reach its destination. The special case for the waiting action is handled separately, which can only occur in a terminal state. 
In order to eliminate the aforementioned intermediate states, the time is progressed by $t((\mathbf{s}, \mathbf{t}),\mathbf{a})$ instead of $1$ within the successor state computation~\eqref{eqn:Tiprime}.  However, after this change, the fact that not all transitions take the same amount of time creates an obstacle in the policy synthesis. In particular, since different transitions can take different amounts of time, the transition durations should be reflected within the cost and the value functions to minimize the average restoration time. Note that $V_n^\pi(\cdot)$~\eqref{def:value} and $V_n^*(\cdot)$~\eqref{def:opvalue} consider the steps in terms of the MDP transitions, not the time required to complete a transition after the proposed modification.

To overcome the discussed issues, a reduced MDP $M' = (S', A', p', c_n)$ together with a modified value function $\mathbb{V}_n^\pi(\cdot)$ is proposed in Definition~\ref{def:modified_mdp}.

\begin{definition}\label{def:modified_mdp} Given an MDP defined as in Section~\ref{sec:model} with cost $c(\cdot)$~\eqref{def:c}, the reduced MDP $M' = (S', A', p', c_n)$ and its value function are defined as follows:
\begin{itemize}
\item $S' \subseteq S$ and $S - S' = \{s \in S \mid A(s) = \{\textbf{C} \} \}$ where $\textbf{C} = [C, \ldots, C] \in A$, 
\item $A'(s) = A(s)$ for each $s \in S'$,
\item $p'(s_a,a,s_b) = p(s_k, \textbf{C}, s_b)$ iff there is a sequence of states $s_a, s_1, \ldots, s_k, s_b \in S$ such that $p(s_a, a, s_1) = 1$, $A(s_i) = \{\textbf{C}\}$ for each $i=1,\ldots,k$, and  $p(s_i, \textbf{C}, s_{i+1}) = 1$ for each $i=1,\ldots,k-1$, 
\item 
 \begin{equation}
\label{def:cn}
    c_n(s,\mathbf{a},s') = c(s,\mathbf{a},s') \times min(n, t(s,\mathbf{a}))
\end{equation}
\end{itemize}
The value function for $M'$ is defined as 
\begin{equation}\label{def:modval}
        \mathbb{V}_n^{\pi}(s) = \sum_{s' \in S} p'(s' \mid s, \pi(s)) ( c_n(s, \pi(s), s') + \mathbb{V}_{n-t(s, \pi(s))}^{\pi}(s') ) \text{ for } n > 0
\end{equation}
and $\mathbb{V}_n^{\pi}(s) = 0$ for all $n \leq 0$. Finally, the modified value $\mathbb{V}_n^{*}(\cdot)$ of the optimal policy  $\pi^*$ (i.e., $\mathbb{V}_n^{\pi^*}(\cdot)$) is 
\begin{equation}\label{def:opmodval}
        \mathbb{V}_n^{*}(s) = \min_{\mathbf{a} \in A} \sum_{s' \in S} p'(s' \mid S, \mathbf{a}) ( c_n(s, \mathbf{a}, s') + \mathbb{V}_{n-t(s,\mathbf{a})}^{*}(s') ) \text{ for } n > 0
\end{equation}
and $\mathbb{V}_n^{*}(s) = 0$ when $n \leq 0$. 
\end{definition}

Intuitively, the reduced MDP is obtained by removing the states that have completely deterministic transitions with the only feasible action of $\textbf{C}$, and $\mathbb{V}_n^\pi(\cdot)$ yields the expected accumulated cost when the policy $\pi$ is followed for $n$ units of time via the new cost formulation $c_n(\cdot)$~\eqref{def:cn} and $\mathbb{V}_n^*(\cdot)$ is its optimal counterpart.
Note that the  function $c_n(s,\mathbf{a},s')$ gives the total cost $c(s,\mathbf{a},s')$ incurred for the minimum of $t(s,\mathbf{a})$ and $n$ time units.

Although we modified the state space of the MDP and the value function, the resulting optimal policy is equivalent to the one from the unaltered MDP and value function formally given in Proposition~\ref{prop:value_eq}. The proof is omitted for space limitations. It is based on the fact that only deterministic transitions are eliminated and that the modified cost function $c_n$ accounts for the accumulated cost between these transitions.

\begin{proposition}\label{prop:value_eq} Let $M = (S, A, p, c)$ be the original MDP with a cost formulation as in~\eqref{def:c} and value function $V_n^{\pi}$, and let $M' = (S', A', p', c_n)$ with $\mathbb{V}_n^{\pi}$ be the modified one as given in Definition~\ref{def:modified_mdp}.
Then for any policy $\pi$ of $M$, its projection $\pi'$ on $M'$ (i.e. $\pi(s) = \pi'(s)$ for any $s \in S'$), $n \geq 0$ and $s_a \in S'$ it holds that 
\begin{align}\label{eq:value_eq}
V_n^{\pi}(s_a)  = \mathbb{V}_n^{\pi}(s_a)
\end{align}
\end{proposition}

Finally, the optimal policy $\pi^\star$ of $M$ can be obtained from the optimal policy $\pi'^\star$ of $M'$ by assigning $\mathcal{C}$ to the states that are removed during the reduction.

\begin{corollary}\label{corol:value_eq}
  Let $M = (S, A, p, c)$ be the original MDP with a cost formulation as in~\eqref{def:c} and value function $V_n^{\pi}$, and let $M' = (S', A', p', c_n)$ with $\mathbb{V}_n^{\pi}$ be the modified one as given in Definition~\ref{def:modified_mdp}. A policy $\pi'^\star$ is optimal for $M'$ if and only if policy $\pi^\star$ is optimal for $M$, where
  \begin{equation} \label{eq:opt_eq}
    \pi^\star(s) = 
    \begin{cases}
      \pi'^\star(s) & \textrm{ if } s \in S' \\
      \mathcal{C}  & \textrm{ otherwise } 
    \end{cases}
  \end{equation}
\end{corollary}
The proof follows from the equality of the value functions~\eqref{eq:value_eq}.

\subsection{Action Elimination}
\label{sub:actelem}

In this section, we propose several rules to reduce the state space of the MDP by determining non-optimal actions during construction.

\subsubsection{Eliminating Non-Optimal Field Team Permutations for the Same Set of Target Buses}
\label{sub:nop} 

The main idea of this optimization strategy is to eliminate non-optimal actions among the ones that assign the same set of new target buses to the same set of field teams. For an action $\mathbf{a} = [a_1, \ldots, a_F] \in A((\mathbf{s}, \mathbf{t}))$, the set of field teams that are assigned to new target buses via $\mathbf{a}$ is denoted as $\mathbf{a}^{teams} = \{ i  \mid  1 \leq i \leq N, a_i \in \busset\}$ and the set of new target buses assigned by $\mathbf{a}$ is denoted as $\mathbf{a}^{buses} = \{ a_i  \mid  1 \leq i \leq N, a_i \in \busset\}$.  
Notice that for any feasible action $\mathbf{a} \in A((\mathbf{s}, \mathbf{a}))$, all actions $\mathbf{a}' = [a'_1, \ldots, a'_F]$ that satisfy the following three conditions ($(i)$, $(ii)$, $(iii)$) are also feasible (see~\eqref{def:AiS}), i.e., all actions $\mathbf{a'}$ that $(i)$ have the same set of teams assigned to new buses as $\mathbf{a}$, $(ii)$ have the same set of target buses as $\mathbf{a}$, and $(iii)$ assign the same action as $\mathbf{a}$ to any team that is not in the set $\mathbf{a}^{teams}$. 
\begin{align*}
(i) \quad & \mathbf{a}^{teams}  = \mathbf{a'}^{teams}, \\
(ii) \quad & \mathbf{a}^{buses} = \mathbf{a'}^{buses}, \\
(iii) \quad & a'_i = a_i \text{ if } i \not \in \mathbf{a}^{teams}
\end{align*}
Each such feasible action represents a different team-bus assignment (i.e., permutation) for the same set of field teams $\mathbf{a}^{teams}$ and the same set of target buses $\mathbf{a}^{buses}$; such actions are referred as \textit{compatible}.
For an action $\mathbf{a}$ and state $(\mathbf{s}, \mathbf{t})$, let us denote the travel time required to reach a bus $i \in \mathbf{a}^{buses}$ by $d(\mathbf{a}, i) = time(j, i)$ where $a_{j} = i$.  Given two compatible actions $\mathbf{a}$ and $\mathbf{a}'$, if one of them, say $\mathbf{a}$, has longer or equal travel durations for all target buses $d(\mathbf{a}, i) \geq d(\mathbf{a}', i), \forall i \in \mathbf{a}^{buses}$, then $\mathbf{a}$ can not be the optimal action in $(\mathbf{s}, \mathbf{t})$. Thus, it can safely be eliminated from the set of feasible actions of  $(\mathbf{s}, \mathbf{t})$. This follows from the fact that all teams are identical apart from their positions, i.e., they travel at the same speed.

We also extend this rule to eliminate the actions that order the teams to exchange buses with each other, e.g., one of the teams moves to the location of another and vice-versa.
Obviously, just ordering these teams to wait is a better option.
Note that such cases can only arise when multiple teams are waiting on buses that are not energizable.

For any subset of teams $\mathbf{x} \subseteq \mathbf{a}^{teams}$, if the set of their target buses is the same as their current locations, then the given action $\mathbf{a}$ is either non-optimal or, at best, equivalent to ordering these teams ($\mathbf{x}$) to wait.
Therefore, the action is eliminated.

To summarize, an action $\mathbf{a} \in A((\mathbf{s}, \mathbf{t}))$ is eliminated if there is another action $\mathbf{a'}  \in A((\mathbf{s}, \mathbf{t}))$ that is compatible with $\mathbf{a}$ and 
\begin{align}
& d(\mathbf{a}, i) \geq d(\mathbf{a}', i), & \forall i \in \mathbf{a}^{buses}         \text{           or          }\\
& \{ t_{i,1} \mid i \in \mathbf{x} \} = \{ a_i \mid i \in \mathbf{x} \}          &      \text{      for some set }   \mathbf{x} \subseteq \mathbf{a}^{teams} \label{def:waitelem}
\end{align}
Note that, in both cases, it is straightforward to show that the action $\textbf{a}'$ leads a value that is not worse than the value obtained with $\textbf{a}$. 

\subsubsection{Wait until an Energization Attempt}
\label{sub:wmt}

The deterministic transitions are assumed to be eliminated with modified value function, as explained in Section~\ref{sub:modval} for this performance optimization. This rule ensures that if a team reaches its destination bus and there have been no energization attempts since it received its last command, it will wait on that bus until an energization attempt occurs in the network.
Intuitively, if this team moves to a new bus without trying to energize the bus, then the policy could have sent it directly to that new bus instead of the one it just arrived, which would be more efficient due to triangle inequality.

Let $s_p=(\mathbf{s}^w,\mathbf{t}^p)$ ($p$ for previous) be an arbitrarily selected state with a successor state $s_w=(\mathbf{s}^w,\mathbf{t}^w)$ ($w$ for waiting) with the same distribution system state $\mathbf{s}^w$, i.e., $p(s_p, a^p, s_w) > 0$ for some $a^p \in A(s_p)$. 
The fact that the distribution system state is the same in both $s_p$ and $s_w$ necessitates the following about the team states $\mathbf{t}^w = [t^w_1, t^w_2, \ldots, t^w_N]$ in $s_w$:
\begin{gather}
t^w_{i,2} = 0 \implies t^w_{i,1} \in \beta_j(\mathbf{s}^w), \text{ for some } j > 1  \text{ for each } i=1,\ldots,N  \\ \nonumber
t^w_{i,2} > 0 \text{ and } t^w_{i,1} \in \beta_1(\mathbf{s}) \text{ for some } i=1,\ldots,N 
\end{gather}
i.e., in $s_w$ all teams that are not en route are on unknown buses that cannot be energized immediately, and there is at least one en route team to an energizable bus $\in \beta_1(\mathbf{s}^w)$.
This follows from the progress condition and the energization rules.
All teams that are not en route must be on unknown buses because if they were on known ($D$ or $E$) buses, then this would mean that an energization attempt has occurred in the last transition, which would prevent $s_w$ and $s_p$ from having the same state.
Similarly, since no team has arrived at a bus in $\beta_1(\mathbf{s}^w)$, there must be at least one en route team to a bus in $\beta_1(\mathbf{s}^w)$ in order to satisfy the progress condition.

This action elimination rule proposes that the only action in $s_w$ should be the deterministic action $a^w$ given in \eqref{eq:aw}, if the prior state is $s_p$.
Thus, we can eliminate $s_w$ completely, update the transition duration of $(s_p, a^p)$, and set the successors of the state-action pair $(s_w, a^w)$ directly to $(s_p, a^p)$.
\begin{equation}
  \label{eq:aw}
  a^w = [a^w_1, a^w_2, \ldots, a^w_N]
  \text{ where } a^w_i = \begin{cases}
    W & \text{if } t^w_{i,2} = 0 \\
    t^w_{i,1} & \text{if } t^w_{i,2} > 0
  \end{cases}
\end{equation}

\begin{proposition}
  \label{prop:waituntil}
  If $a^p$ is optimal in $s_p$, then $a^w$ is also optimal in $s_w$.
\end{proposition}
\begin{proof}
  Assume for the sake of contradiction that $a^w$ is not optimal in $s_w$.
  Instead, there exists another action $a^o$ that is optimal in $s_w$.
  Taking this action in $s_w$ will lead to a set of states $S^f$.
  Note that for all $(\mathbf{s}, \mathbf{t}) \in S^f$, team states $\mathbf{t}$ are the same.

  We can build an action $a^c = [a^c_1, a^c_2, \ldots, a^c_N]$ for $s_p$ such that
  \begin{equation}
    \label{eq:prop:aci}
    a^c_i = \begin{cases}
      a^o_i & \text{if } t^p_{i,2} = 0 \text{ and } t^w_{i,2} = 0 \\
      t^w_{i,1}  \ ( =a^p_i) & \text{if } t^p_{i,2} = 0 \text{ and } t^w_{i,2} = 0 \\
      t^p_{i,1} & \text{if } t^p_{i,2} > 0
    \end{cases}
  \end{equation}
  Essentially, $a^c$ sends the teams to the same destinations as $a^o$ but from state $s_p$.
  Consequently, taking $a^c$ in $s_p$ will lead to the same set of successor states $S^f$.
  By triangle inequality, reaching $S^f$ through $s_w$ must not be any faster than the direct path using $a^c$.  Therefore, if $a^o$ is optimal in $s_w$, $a^c$ must also be optimal in $s_p$, which avoids $s_w$ altogether.
  This leads to a contradiction since the proposition assumes that $a^p$ is optimal in $s_p$.
\end{proof}

This action elimination method is integrated into the MDP model by progressing the time until a team arrives at a bus in $\beta_1(\mathbf{s}^w)$ as in~\eqref{eqn:tSaWupdate} instead of progressing the time until a team arrives at its destination~\eqref{eqn:tSaW} during the reduced model construction (Definition~\ref{def:modified_mdp}).
 \begin{align}
    t'((\mathbf{s}, \mathbf{t}),\mathbf{a}) & = \min_{1 \leq i \leq F, a_i \neq W, a_i \in \beta_1(\mathbf{s})} rem(\mathbf{t},\mathbf{a},i) \quad\textrm{ when } \mathbf{a} \neq \mathbf{W}  \nonumber \\
    t'((\mathbf{s}, \mathbf{t}),\mathbf{a}) & = 1 \quad \textrm{ when } \mathbf{a} = \mathbf{W}
\label{eqn:tSaWupdate}
\end{align}

\subsubsection{Components on the Way}
\label{sub:components_on_the_way}

This rule eliminates actions that send a team from $a$ to $b$ when another bus energizable $c$ with $time(a,c) + time(c,b) = time(a,b)$ is available. In other words, $c$ is a bus on the way from $a$ to $b$, and therefore, it can be visited with no additional cost.
To formalize this rule, assume that the given state is $s_p = (\mathbf{s}^p, \mathbf{t}^p)$, and consider a team $x$ (with ${t}^p_x \in \busset$) and 2 actions $a^1$ and $a^2$ that differ only by the command to team $x$, i.e., $a^1_i = a^2_i$ for all $i \neq x$ and $a^1_x \neq a^2_x$.
If $time({t}^p_x, a^1_x) = time({t}^p_x, a^2_x) + time(a^2_x, a^1_x)$, then we can safely eliminate $a^1$ from $A(s_p)$.
Because even if $a^1$ is an optimal action, $a^2$ has to be optimal as well since the team $x$ can reach $a^1_x$ after reaching $a^2_x$ at the same time.
Hence, the cost of $a^1$ is greater or equal to the cost of $a^2$. Note that it is never optimal to delay the energization of a bus. In particular, the cost incurred by a bus $i$ for optimization horizon $h$ is $P_f(i) \cdot h$ if it is attempted to be energized immediately.
Delaying the energization for $k > 0$ units of time would make this $k + P_f(i) \cdot (h-k)$, which is necessarily larger than $P_f(i) \cdot h$ for $P_f(i) < 1$. Thus, due to the given time condition, even if  $a^2_x$ is not energizable in an $a^2$ successor of $s_p$, $a_2$ is also optimal if $a_1$ is optimal. This rule eliminates action $a_2$ and the corresponding successor states during MDP construction instead of the policy synthesis.

Note that optimal actions can be eliminated if this rule and ``Wait until an Energization Attempt" are used together. In particular, this rule enforces a team to visit a bus $c$ if that is on the way to another unknown bus $b$ (i.e., no additional time to reach $j$), and the other rule enforces a team to wait on bus $c$ if it is in unknown status. Thus, these two optimization rules are not compatible. In order to ensure that they can be used together without sacrificing optimality, the component on the way must be in $\beta_1(s_p)$, i.e., energizable.

\subsection{Equivalent Team State Permutations}
\label{sub:teamperm}

All teams are assumed to be equivalent, i.e., all energization attempts that team-$i$ can perform can also be performed by team-$j$. Consequently, two team states $\mathbf{t}^a =[t^a_1, \ldots, t^a_F]$ and $\mathbf{t}^b =[t^b_1, \ldots, t^b_F]$ that differ only by the locations of the two teams $i$ and $j$, i.e., $t^a_k =  t^b_k$ if $k \not \in \{i,j\}$, $t^a_i = t^b_j$ and $t^a_j = t^b_i$, are the same in terms of the restoration purposes. In particular, for any successor of $\mathbf{t}^a$, say ${\mathbf{t}^a}' = [{t^a_1}', \ldots, {t^a_F}']$, there exists a successor of $\mathbf{t}^b$, ${\mathbf{t}^b}' =[{t^b_1}', \ldots, {t^b_F}']$, such that ${t^a_k}' =  {t^b_k}'$ if $k \not \in \{i,j\}$, ${t^a_i}' = {t^b_j}'$ and ${t^a_j}' = {t^b_i}'$. Generalizing this, for a given state $\mathbf{t} = [t_1, \ldots, t_F]$, we can define the set of equivalent team states as follows:
\begin{equation}\label{eq:equivalence_TF}
 [\![\mathbf{t}]\!] =  \{ [t_1', \ldots, t_F'] \mid \forall i \left( \exists j, t_i' = t_j \right) \text{ and } \forall i \left( \exists j, t_i = t_j' \right) \}
\end{equation}
The equivalence classes for team states~\eqref{eq:equivalence_TF} induce equivalence classes over the MDP states, i.e., $ [\![( \mathbf{s}, \mathbf{t}) ]\!]  =     \{ (\mathbf{s}, \mathbf{t}') \mid   \mathbf{t}' \in  [\![\mathbf{t}]\!] \}$.  Among these equivalent states, it is sufficient to explore only one without loss of generality. This idea is integrated into the developed tool by using a lexicographical ordering over the team locations.

%% file: sections/experiment.tex
\section{Experimental Evaluation}
\label{sec:experiment}

We implemented the proposed MDP-based restoration strategy synthesis method and the performance optimization approaches in a tool called \textbf{PowerRAFT: Power Restoration Application with Field Teams} 
using Rust and evaluated them over several cases. 
 All experiments were run on a desktop computer with Intel i9-10900X 4.50 GHz CPU and 48 GB of RAM, running Ubuntu 22.04.1 LTS. 
 
The experimental results are reported in the following sections. First, Section~\ref{sub:samplesys} introduces the sample systems used in the experiments. Then, Section~\ref{sec:exp_performance} presents the results of the experiments that are performed to evaluate the effects of the proposed optimization techniques over the model size and the computation time. The relation of the system parameters (the number of buses, branches, and field teams) with the expected restoration time is analyzed in Section~\ref{sec:exp_system_parameters}. Finally, the partitioning-based approach and experiments over IEEE-37 and IEEE-123 systems are given in Section~\ref{sec:partition}.

\begin{figure}[h]
  \centering
  \centerline{\includegraphics[width=0.9\textwidth]{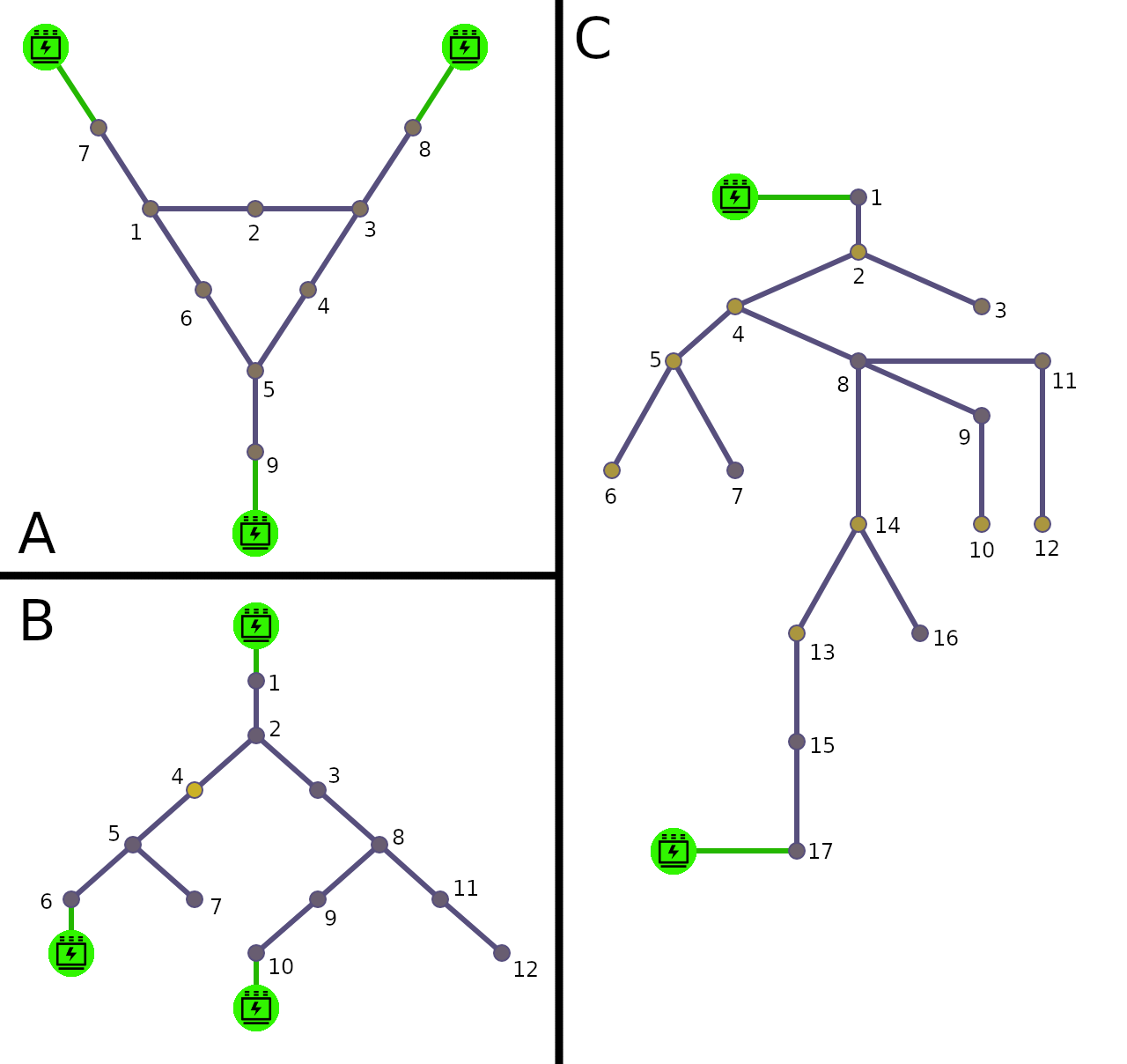}}
  \caption{The distribution systems used in the experiments: WSCC 9-bus (A), 12-bus (B), real-life 17-bus (C).}
  \label{fig:sample_systems}
\end{figure}

\subsection{Sample Systems}
\label{sub:samplesys}

Three sample systems are used in the experiments:  WSCC 9-bus distribution system \citep{powerflowrepo}, a 12-bus distribution system~\citep{Arpali2019}, and a real-life 17-bus distribution system. All three systems are shown in Figure~\ref{fig:sample_systems}. In 9-bus and 12-bus systems, as-the-crow-flies distance between buses is used to compute the travel times. The largest distance between neighboring buses is used as the divisor in discretization so that the travel time for each pair of connected buses is always 1, i.e, $time(i,j)  = 1$ for each $(i,j) \in \busedges$. 
The network topology of the 17-bus system is based on a real-life distribution system with fictitious locations to preserve confidential information.
In this system, the average (not the largest as in the previous systems) distance between neighboring buses is used as the divisor in the discretization.
The $P_f$ values used in the experiments are given in Table~\ref{table:Pf}.
  Since these experiments are synthetic, these $P_f$ values were assigned arbitrarily as opposed to using fragility curves described in Section~\ref{sub:Pf}.
  Nevertheless, our MDP model generates exact solutions regardless of the $P_f$ values.

\begin{table}[h]
\centering
\begin{tabular}{|c|l|}
\hline
9-bus & $P_f(i) = 0.25 $ for each $i \in \{1,\ldots,9\}$ \\ \hline
12-bus & $P_f(i) = 0.1 $ for each $i \in \{1,2,3\} \cup \{5,\dots, 12\}$,  $P_f(4) = 0.7 $\\ \hline
17-bus & \begin{tabular}[l]{@{}l@{}}$P_f(1) = 0.125$, $P_f(2) = 0.5$, $P_f(3) = 0.25$, $P_f(4)  = 0.5$, $P_f(5) = 0.5$, \\ $P_f(6)  = 0.5$, $P_f(7) = 0.125$, $P_f(8) = 0.125$, $P_f(9)  = 0.125$, \\ $P_f(10) = 0.5$, $P_f(11) = 0.25$, $P_f(12) = 0.5$, $P_f(13) = 0.5$,  \\ $P_f(14)  = 0.5$, $P_f(15) = 0.125$, $P_f(16) = 0.125$, $P_f(17) = 0.125$
\end{tabular}
\\ \hline 
\end{tabular}
\caption{The probability of failure values used for the sample systems.}
\label{table:Pf}
\end{table}

\begin{table}[]
\centering
\begin{tabular}{|l|l|l|}
\hline
V & Model reduction via modified \underline{v}alue function & Section~\ref{sub:modval} \\ \hline
P &   Non-optimal \underline{p}ermutations                                          &  Section~\ref{sub:nop}   \\ \hline
W & \underline{W}ait until an energization attempt                                  &   Section~\ref{sub:wmt}    \\ \hline
O &            Components \underline{o}n the way                                 &      Section~\ref{sub:components_on_the_way}   \\ \hline
S &                  Equivalent team  \underline{s}tate permutations                          & Section~\ref{sub:teamperm}      \\ \hline

\end{tabular}
\caption{The abbreviations used for the performance optimization methods.}
\label{table:abbr}
\end{table}

\subsection{Performance Evaluation for the Optimizations}\label{sec:exp_performance}

This section presents experiments that are conducted using various combinations of action elimination methods and the modified value function approach, aimed at evaluating their impact on reducing the model size and computation time.
The abbreviations used for these methods are given in Table~\ref{table:abbr}\footnote{Since W requires V, V was omitted in the labels if W is enabled for the sake of brevity.}.

\begin{table}[h]
  \centering
  \begin{tabular}{|l|r|r|r|r|r|r|}
    \hline
     & \multicolumn{3}{p{10em}|}{9-bus system with starting teams $(9,9,9)$} & \multicolumn{3}{p{10em}|}{ 9-bus system with starting teams $(9,9)$} \\
     \hline
        Optimizations & $t^{MDP}$ & $t^{total}$ & \texttt{\#states} & $t^{MDP}$ & $t^{total}$ & \texttt{\#states} \\
    \hline 
            - &         14.59  &         19.65  &         1770349  &         0.36  &         0.48  &         104588  \\
            V &         14.94  &         21.26  &         1582075  &         0.33  &         0.46  &          83777  \\
            W &         12.76  &         17.72  &         1253179  &         0.29  &         0.39  &          72339  \\
            P &         12.19  &         16.02  &         1527644  &         0.37  &         0.47  &          99365  \\
            O &          1.40  &          1.79  &          267659  &         0.07  &         0.08  &          29408  \\
    P + O + V &          1.27  &          1.64  &          234444  &         0.08  &         0.09  &          25694  \\
    P + O + W &          1.03  &          1.33  &          200514  &         0.06  &         0.07  &          22171  \\
            S &          2.84  &          3.74  &          353559  &         0.18  &         0.22  &          56114  \\
    S + P + V &          2.61  &          3.50  &          300101  &         0.19  &         0.23  &          44142  \\
    S + P + W &          2.10  &          2.78  &          240373  &         0.17  &         0.20  &          38303  \\
    S + O + V & \textbf{ 0.30} & \textbf{ 0.37} & \textbf{  56820} & \textbf{0.04} & \textbf{0.04} &          14317  \\
    S + O + W &          0.41  &          0.51  &           67099  &         0.04  &         0.05  &          14175  \\
    S + P + O &          0.44  &          0.51  &           80392  &         0.06  &         0.07  &          19406  \\
S + P + O + V &          0.43  &          0.52  &           75575  &         0.05  &         0.06  &          16972  \\
S + P + O + W &          0.37  &          0.44  &           64287  &         0.04  &         0.05  & \textbf{ 13997} \\
    \hline 
  \end{tabular}
  \caption{Optimization benchmark results on the 9-bus system, where $t^{MDP}$ is the time elapsed for constructing the MDP (in seconds), $t^{total}$ is the total execution time (in seconds), \texttt{\#states} is number of states in the MDP.}
  \label{table:optbenchmark9}
\end{table}

\begin{table}[h]
  \centering
  \begin{tabular}{|l|r|r|r|r|r|r|}
    \hline
     & \multicolumn{3}{p{12em}|}{12-bus with only bus 1 connected to a substation and starting teams $(1,1,1)$} & \multicolumn{3}{p{12em}|}{12-bus with only buses 1 and 10 connected to substations and starting teams $(1,1)$} \\
     \hline
        Optimizations & $t^{MDP}$ & $t^{total}$ & \texttt{\#states} & $t^{MDP}$ & $t^{total}$ & \texttt{\#states} \\
    \hline 
            - &         15.32  &         25.29  &         2852126  &         0.93  &         1.69  &         296342  \\
            V &         15.57  &         28.53  &         1951947  &         0.74  &         1.51  &         172011  \\
            W &         12.21  &         20.76  &         1392987  &         0.65  &         1.23  &         144691  \\
            P &         15.12  &         23.83  &         2645762  &         1.03  &         1.69  &         287760  \\
            O &          3.35  &          4.92  &          699465  &         0.32  &         0.49  &         116937  \\
    P + O + V &          3.18  &          4.76  &          515816  &         0.32  &         0.46  &          80412  \\
    P + O + W &          2.90  &          4.32  &          501882  &         0.28  &         0.39  &          75062  \\
            S &          2.69  &          4.29  &          517862  &         0.44  &         0.77  &         152235  \\
    S + P + V &          2.57  &          4.26  &          341489  &         0.42  &         0.70  &          87483  \\
    S + P + W &          1.91  &          2.97  &          240839  &         0.36  &         0.55  &          73271  \\
    S + O + V & \textbf{ 0.59} &          0.88  &          109436  & \textbf{0.14} &         0.19  &          42394  \\
    S + O + W &          0.69  &          1.01  &          121026  &         0.14  & \textbf{0.19} &          42655  \\
    S + P + O &          0.72  &          0.93  &          130736  &         0.21  &         0.27  &          63052  \\
S + P + O + V &          0.67  &          0.91  & \textbf{ 106973} &         0.18  &         0.23  &          44262  \\
S + P + O + W &          0.67  &          0.92  &          113014  &         0.16  &         0.20  & \textbf{ 41787} \\
    \hline 
  \end{tabular}
  \caption{Optimization benchmark results on the 12-bus system, where $t^{MDP}$, $t^{total}$  and \texttt{\#states} are defined as in Table~\ref{table:optbenchmark9}.}
  \label{table:optbenchmark12}
\end{table}

\begin{figure}
  \centering
  \includegraphics[width=\textwidth]{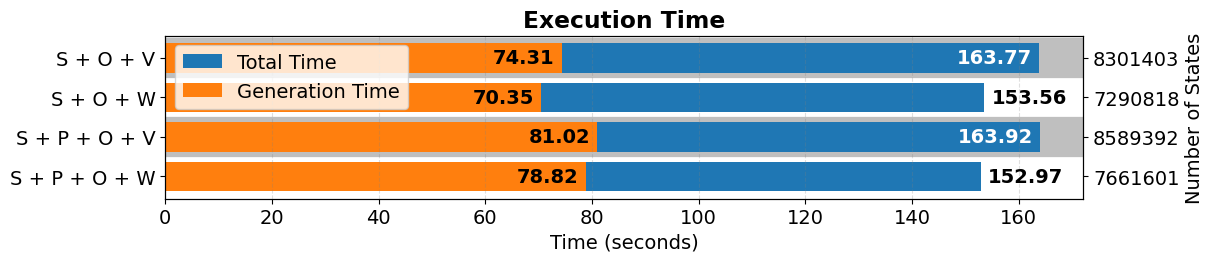}
  \caption{Optimization benchmark results for the 17-bus system with teams starting at $(1,1)$. The MDP construction time and the time are given on the orange and blue bars, respectively. The number of states is given on the right-hand side.}
  \label{fig:opt17}
\end{figure}

The first experiment is conducted on the 9-bus distribution system with 3 or 2 teams starting at bus-9, i.e., the initial state of the teams is $(9, 9, 9)$ or $(9, 9)$.
The second experiment is conducted on the 12-bus system with only bus 1 connected to a substation and teams initially at $(1,1,1)$, and with buses 1 and 10 connected to substations and teams initially at $(1,1)$.
All possible combinations of optimizations are tested, and highlights are reported in Tables \ref{table:optbenchmark9} and \ref{table:optbenchmark12}. We run the best optimization combinations on the larger 17-bus system with $(1,1)$ as the initial state of the teams and plotted the results in Figure \ref{fig:opt17}.

The reported model sizes and the computation times reveal that the optimizations are indispensable for real-life applications since the naive execution time is up to 117 times larger than the fully optimized execution time, and its state count is 91 times larger (12-bus system, first experiment). Furthermore, the reduction ratio increases as the system becomes more complex. Regarding the effects of each optimization method, the components on the way (O) optimization is the most successful one by itself, both in the execution time and the state space reduction. Even though each optimization method is quite successful by itself, the total reduction in the computation time (or the model size) is, in general, less than the sum of the individual rates when the optimization methods are applied together. 
The elimination of the same actions by different optimization methods and the computational overhead of these methods are the main reasons. Nevertheless, applying several optimization techniques together yields better results. 
S+O+V is the best combination in smaller systems (9-bus and 12-bus), whereas S+P+O+W outperforms other combinations in larger 17-bus system. 

\subsection{System Topology and Restoration Time}\label{sec:exp_system_parameters}

\begin{figure}
  \centering
  \includegraphics[width=\textwidth]{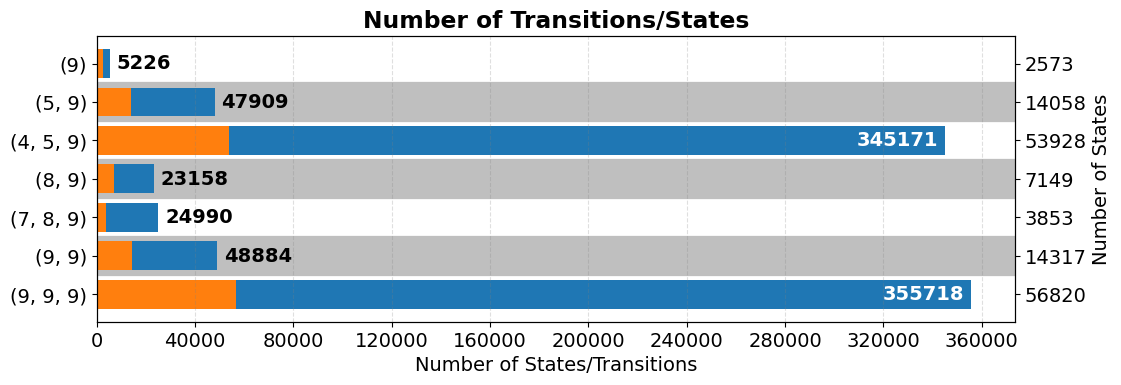}
  \caption{The number of states and transitions for different team configurations in 9-bus system. The orange bar represents the state count, and the actual number of states is given in the right of the plot. The blue bar represents the number of transitions, and the actual number is also given on the plot.}
  \label{fig:teamst}
\end{figure}

\begin{figure}
  \centering
  \includegraphics[width=\textwidth]{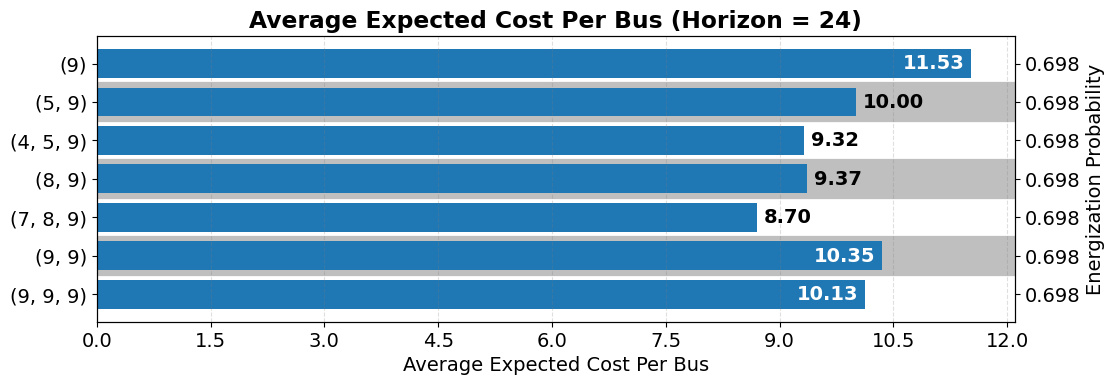}
  \caption{Average expected cost per bus for different team configurations in the 9-bus system system. The initial location of the teams is given on the left, and the average probability of energizing a bus is given on the right.}
  \label{fig:teamac}
\end{figure}

A sequence of experiments is conducted to analyze how the restoration time is affected by the number of teams, branches, and substation connections. The corresponding subsections provide detailed information about these experiments.

\subsubsection{The Number of Teams}

In this experiment, the restoration problem on the 9-bus system given in Figure~\ref{fig:sample_systems} is solved with different team configurations in order to analyze the impact of the number of teams on the restoration time and model size.
These experiments are run using S+O+V optimizations, which yield the smallest MDP size on this system as shown in Table~\ref{table:optbenchmark9}.

The number of states and transitions for different team configurations are given in Figure~\ref{fig:teamst}. The ratio between the number of transitions and the number of states increases with the team count.   The MDP size, in general, is exponential in the number of states. An exception to this is observed when the teams are located on buses connected to an energy source, e.g., $(9) \to (8, 9) \to (7, 8, 9)$. In this case, since the teams immediately attempt to energize the buses at which they are located, many states are eliminated.

\begin{figure}
  \centering
  \includegraphics[width=\textwidth]{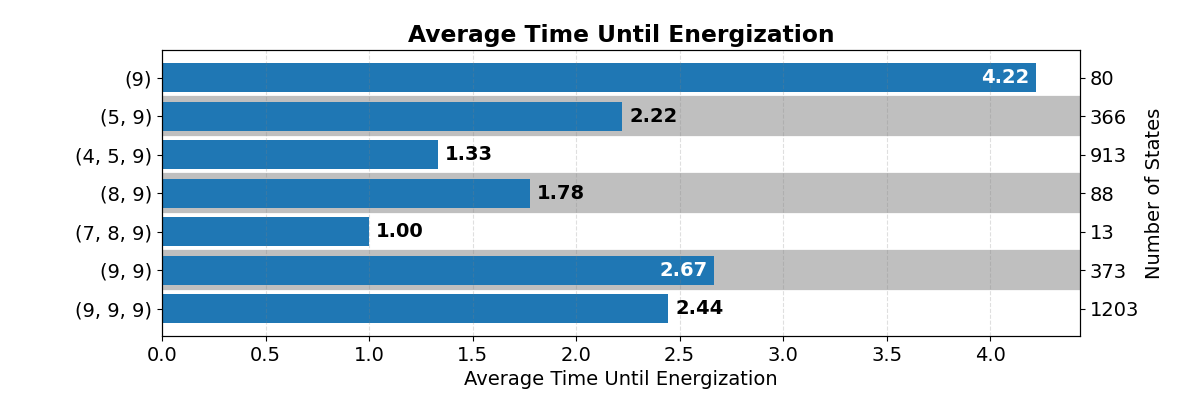}
  \caption{Average time to energize a bus for different team configurations in the 9-bus system when $P_f(i) = 0$ for each bus-$i$.  The initial location of the teams is given on the left, and the number of states of the MDP model is given on the right.}
  \label{fig:teamavg}
\end{figure}

The average expected cost per bus for the initial state $s_0$, i.e., $\mathbb{V}_{24}^{*}(s_0) / 9$,  is given in Figure \ref{fig:teamac} for optimization horizon 24.
In addition, the average time to energize a bus ($\mathbb{V}_{30}^{*}(s_0) / 9$) when no bus is damaged is given in Figure \ref{fig:teamavg} ($P_f(i) = 0$ for each bus-$i$). 
Note that $\mathbb{V}_{n}^{*}(s_0) / N$ does not yield the average time to energize a bus when $P_f(i) > 0$ for some bus-$i$ since there are reachable terminal states in which a bus is not energized.
The expression $\mathbb{V}_{n}^{*}(s_0)$ equals the sum of the cost incurred along the path from $s_0$ to each terminal state $s_t$ multiplied by its probability.
When $P_f(i)= 0$ for each bus-$i$, all buses are energized in all reachable terminal states; hence, all self-transitions in terminal states have 0 costs.

Adding a new team reduces the energization time significantly. Furthermore, the geographic distribution of the teams has a considerable effect on the energization time. First, it is preferable to have the teams away from each other, i.e., $(9) \to (5, 9) \to (4, 5, 9)$ is better than $(9) \to (9, 9) \to (9, 9, 9)$ since additional teams can easily travel to the other buses. Second, locating the teams close to the sources is preferable, i.e., $(9) \to (8, 9) \to (7, 8, 9)$ is better than $(9) \to (5, 9) \to (4, 5, 9)$, since the bus on which the new team is located can be energized immediately.  In the case of $(9) \to (9, 9) \to (9, 9, 9)$, adding the third team provided minimal benefit. Therefore, there are diminishing returns to adding a new team to the same bus.

Finally, note that the state counts in Figure~\ref{fig:teamavg} are much smaller than the previous case reported in Table~\ref{table:optbenchmark9}.
This is because the implementation discards all transitions with 0 probability, thereby avoiding the exploration of unreachable successor states.
In real-life applications, this fact can be easily exploited by rounding the $P_f$ values of some buses to $0$ or $1$ to compute a simpler, approximate solution. %

\subsubsection{The Number of Branches}

\begin{figure}
  \centering
  \includegraphics[width=\textwidth]{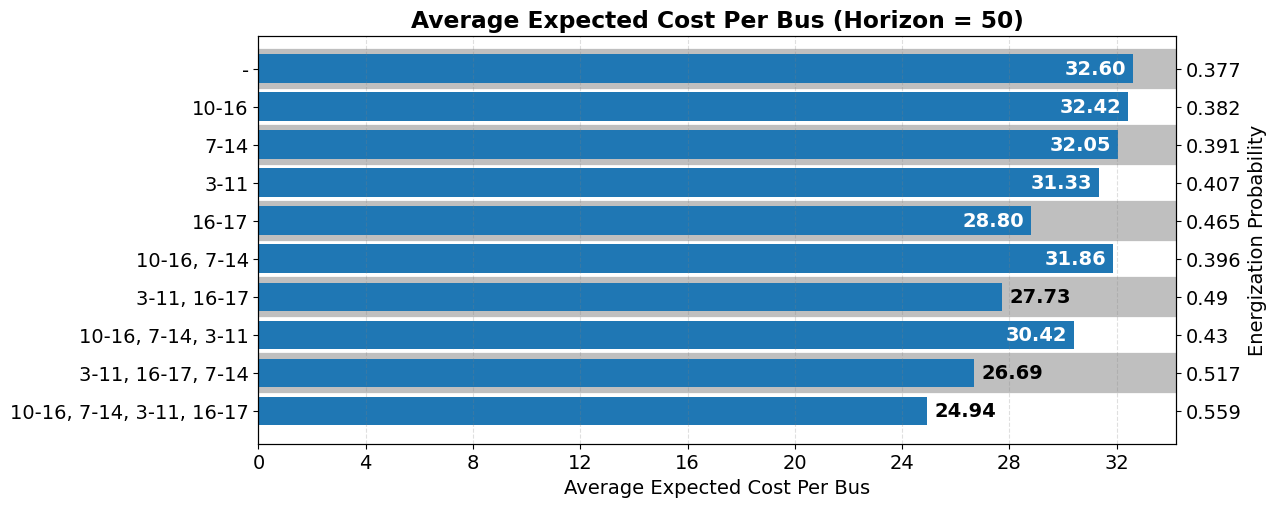}
  \caption{The set of additional branches is shown on the left, where a branch between buses $i$ and $j$ is shown with $i-j$. The average expected cost per bus is shown with the blue bars, and the numbers are given on the plot. The average probability of energizing a bus is given on the right.}
  \label{fig:bavg}
\end{figure}

Redundant branches are used to improve the robustness of power grids~\cite{gonen2015electric}, which can be particularly beneficial in post-disaster scenarios. 
In this experiment, new branches are added to a modified version of the 17-bus system to assess the impact of additional branches on restoration time. 
Specifically,  buses 6 and 12 are removed, the teams start at buses 1 and 16 on the remaining 15-bus system, and the model is constructed using the S+P+O+W optimization method. 
Figure~\ref{fig:bavg} shows the average expected cost per bus for various scenarios with additional branches.

The experiments show that a redundant branch always reduces the cost. The reduction rate is higher when the new branch connects a bus that is close to an energy source to a bus that is not. For example,  $(3, 11)$ and $(16, 17)$ are better than $(7, 14)$ and $(10, 16)$, both individually and together. 

\subsubsection{The Number of Substation Connections}


\begin{figure}
  \centering
  \includegraphics[width=\textwidth]{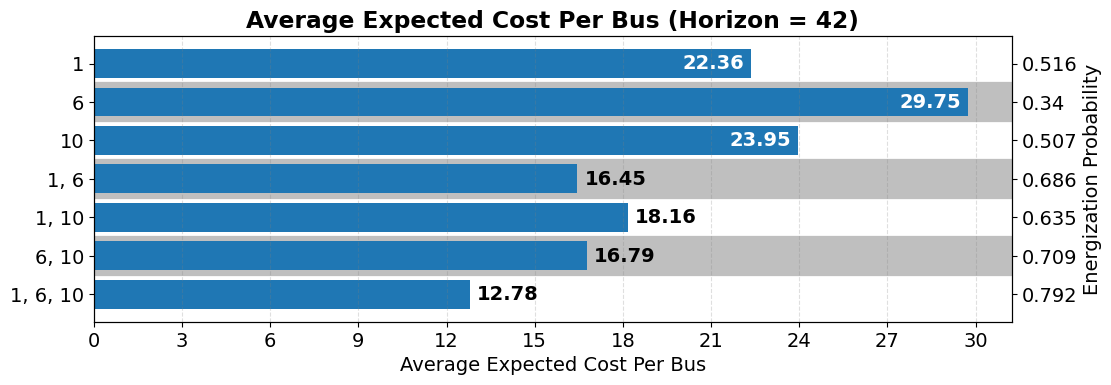}
  \caption{The set of buses connected to substations is shown on the left. The average expected cost per bus is shown with the blue bars, and the numbers are given on the plot. The average probability of energizing a bus is given on the right.
  }
  \label{fig:tgavg}
\end{figure}

The robustness of the distribution system can also be improved by connecting more buses to substations. In these experiments, various substation connection scenarios are analyzed for the restoration problem using the 12-bus system (Figure~\ref{fig:sample_systems}).  In these scenarios, two field teams start at bus $1$, and the S+P+O+W method is used for model construction. The results are reported in Figure~\ref{fig:tgavg}. 

As given in Table~\ref{table:Pf}, bus-4 has a very high $P_f$ value. The network becomes disconnected when this bus is damaged due to the lack of redundant branches in this system. Thus, it is more advantageous to connect both sides when two connections are possible (e.g., 1,6 and 6,10 are better than 1,10). On the other hand, when the second substation connection is not possible, it is better to connect the larger part (e.g., 1 or 10). 

\subsection{Evaluation of Scalability with Partitioning}\label{sec:partition}

Despite the developed action elimination methods, it is not feasible to apply the proposed method to large distribution systems due to the exponential relation between the system size and the MDP size. To overcome this challenge, we take inspiration from~\cite{preweather,back7} and employ an intuitive partition-based method for large systems.
In this approach, the system is divided into partitions determined by the user, and each partition is handled independently as a separate system. This method yields sub-optimal solutions as the branches connecting the partitions are not used. However, by dividing the network at the minimum cuts, the impact on optimality can be reduced.

The partitioning based experiments are run over IEEE-37 and IEEE-123 systems~\cite{IEEE37}.
To determine the distances between the buses in the IEEE-37 system, the cable lengths are divided by 200 feet, and the result is rounded to the nearest integer. 
On the other hand, buses in the IEEE-123 system are placed arbitrarily while preserving the network topology.

In the IEEE-37 system, to simulate an earthquake in the southern region, higher $P_f$ values are assigned to the buses in the south region. 
The northernmost 8 buses have $P_f = 0$, whereas the southernmost 2 buses have $P_f = 1$.
In the IEEE-123 system, most of the buses have $P_f = 0.25$ and 31 buses $P_f = 0$ since not all areas are affected by the earthquake in such a large system.

\begin{table}[tb]
  \centering
  \begin{tabular}{|l|l|r|r|r|r|}
    \hline
    System & Partition & \texttt{\#buses}& $t^{MDP}$ & $t^{total}$ & \texttt{\#states} \\
    \hline 
    IEEE-37 & Full System & 35 & \textbf{329.10} & \textbf{2475.08} & \textbf{38761504} \\
    \hline 
    \multirow{2}{*}{IEEE-37}
    &    Top & 15 &         0.02  &         0.03  &          2409  \\
    & Bottom & 20 & \textbf{0.11} & \textbf{0.18} & \textbf{31567} \\
    \hline 
    \multirow{5}{*}{IEEE-123}
    & Part A & 28 &          14.02  &           41.66  &          2626049  \\
    & Part B & 32 & \textbf{398.48} & \textbf{1235.58} & \textbf{52869842} \\
    & Part C & 20 &           0.02  &            0.03  &             6498  \\
    & Part D & 22 &           0.30  &            0.52  &            79207  \\
    & Part E & 21 &           0.26  &            0.51  &            75361  \\
    \hline 
  \end{tabular}
  \caption{Partitioning benchmark results. \texttt{\#buses} is the number of buses in the partition, $t^{MDP}$ is the time elapsed for constructing the MDP (in seconds), $t^{total}$ is the total execution time (in seconds), and \texttt{\#states} is the number of states in the MDP. Maximum $t^{MDP}$, $t^{MDP}$, and \texttt{\#states} are printed in bold for each separate case.}
  \label{table:partition}
\end{table}

The benchmark results are given in Table~\ref{table:partition}.
In all cases, only one team is assigned to each partition, and each team starts at the bus through which the system is connected to the substation.
IEEE-37 is the largest system our tool can solve with one team without partitioning. However, partitioning allows us to solve this system with two teams at a fraction of the cost.
The IEEE-123 system is solved by dividing it into 5 partitions, with the largest partition taking disproportionately the longest amount of time due to the exponential increase in the number of states.
If Part B were any larger, it would be necessary to divide it into two regions.

These results suggest that for large systems where the optimizations presented in Section~\ref{sec:opt} are not sufficient, partitioning is a viable way to generate sub-optimal energization strategies.

%% file: sections/conclusion.tex
\section{Conclusion}
\label{sec:conclusion}

This paper presents an MDP-based solution for optimizing the coordination of field teams to energize an electrical distribution system hit by an earthquake. The proposed method builds upon previous work \citep{Arpali2019}, which assumed a remote control system rather than incorporating field teams and their travel times into the model.
In the proposed solution, the MDP model integrates field teams and their travel times to synthesize a policy that minimizes the expected restoration time. To cope with the increased MDP size, the paper also develops several action elimination methods that determine non-optimal actions during MDP construction and eliminate them. In addition to the indispensability of the elimination methods with up to 99 percent reduction rate in computation time, our experiments indicate the need for other techniques tailored to larger systems. The partition-based method inspired by~\cite{preweather,back7} allows us to generate suboptimal strategies for considerably larger systems. This strategic approach proved instrumental in tackling challenges posed by the model's exponential dependence on system size. Moreover, our experimental analysis aligns with the robustness literature, confirming that redundancy in system design significantly contributes to reduced energization times in post-disaster scenarios.  Furthermore, the experiments provide valuable insights into the team localization dynamics.

Future research could extend this work to incorporate stochastic travel times, as in \cite{Khani2018RealtimeTC}. Additionally, the proposed method could be further optimized to handle much larger inputs without partitioning, e.g., by employing deep reinforcement learning. Finally, the model could be extended to incorporate distributed energy resources as in \citep{Arpali2019} or power-flow analysis as in~\cite{Yilmaz2023}. 

\section*{Acknowledgements}

Funding: This work is supported by Scientific and Technological Research Council of Turkey (TUBITAK) under project number 118E183.

%% file: sections/proposition-proof.tex
\section{Proof of Proposition \ref{prop:value_eq}}


\begin{proof}
Let $s_a$ be an arbitrary state in $S'$, and $s_a, s_1, \ldots, s_k$ with $k \geq 0$ be a sequence of deterministic transitions initiated with $\pi(s_a)$ ($= \pi'(s_a)$ by definition of $\pi'$) as given in the second point of Definition 4.1, 
i.e.,  $p(s_a, \pi(s_a), s_1) = 1$ and $A(s_i) = \{\textbf{C}\}$ for each $i=1,\ldots,k$. 
  Finally, let $c_{s_a} = c(s_a, \pi(s_a), s_1)$. \footnote{Note that $c_{s_a} = c(s_a, \pi(s_a), s_b)$ for some $s_b \in S'$ when $k=0$.}
  Observe that by (27)
\begin{equation}\label{eq:cost_equality}
c(s_i, \pi(s_i), \cdot) = c_{s_a}  \text{      for      }i=1,\ldots,k 
\end{equation}
since no energization attempt is possible at these states. 
The equivalence property (32) 
is proven by induction on $n$. \\
\textit{Base-case.} The equality trivially holds for $n=0$ by the definitions of $V_0(\cdot)$ (1) 
 and $\mathbb{V}_0(\cdot)$ (31). 
Furthermore, consider the case when the time to complete traveling for a team in $s_a$ is greater than the value horizon, i.e, $k+1 = t(s_a, \pi(s_a)) > n$. The value function for $s_a$ and $\pi$ is given in~\eqref{eq:proof_value_def}. Since the first $k$ transitions are deterministic ($p(s_1 \mid s_a, \pi(s_a)) = 1)$, the equation simplifies to~\eqref{eq:proof_value2}, where the summation is also unfolded for $V_{n-1}^\pi(s_1)$. Via the iterative application of this approach, we obtain  $n \cdot c_{s_a}$ as the value of $s_a$ under policy $\pi$ for horizon $n$~\eqref{eq:proof_value3}.
\begin{align}
V_n^{\pi}(s_a) & = \sum_{s' \in S} p(s' \mid s_a, \pi(s_a)) ( c(s_a, \pi(s_a), s') + V_{n-1}^{\pi}(s') ) \label{eq:proof_value_def} \\
		       & = 1 \cdot c_{s_a} + \sum_{s' \in S} p(s' \mid s_1, \mathcal{C}) ( c(s_1, \pi(s_1), s') + V_{n-2}^{\pi}(s') ) \label{eq:proof_value2}\\
		       & = n \cdot c_{s_a} +  V_{n-n}^{\pi}(s_n)\label{eq:proof_value3}
\end{align}
Similarly, the modified value function for $s_a$ and policy $\pi'$ is given in~\eqref{eq:proof_value_modified_def}. 
As $min(n, t(s_a, \pi'(s_a)))$ was assumed to be $n$ and $\mathbb{V}_{n-k-1}^{\pi}(s_b)$ is 0 (since $n < k+1$), we also obtain $c_{s_a} \cdot n $ as the value of $s_a$ for horizon $n$ for $M'$~\eqref{eq:proof_value_modified2}, which concludes the analysis for the base case.
\begin{align}
\mathbb{V}_n^{\pi}(s_a) & =  \sum_{s' \in S'} p'(s' \mid s_a, \pi'(s_a)) ( c_n(s_a, \pi'(s_a), s') + \mathbb{V}_{n-t(s_a, \pi(s_a))}^{\pi}(s') )  \label{eq:proof_value_modified_def} \\
& = c_{s_a} \cdot n + \mathbb{V}_{n-k-1}^{\pi}(s_b)  \label{eq:proof_value_modified2} 
\end{align}
\textit{Induction step.} Assume that the property (32) 
holds for all $i < n$. Let $k+1 = t(s_a, \pi(s_a)) \geq n$ (the other case is proven above). Due to the assumption that the first $k$ transitions are deterministic when $\pi(s_a)$ is applied at $s_a$ and the states costs are the same~\eqref{eq:cost_equality}, we reach at~\eqref{eq:is1}. Since $\sum_{s' \in S} p(s' \mid s_k, \mathcal{C}) = 1$, \eqref{eq:is1} is rewritten as \eqref{eq:is2}.
\begin{align}
V_n^{\pi}(s_a) &= k \cdot c_{s_a} + \sum_{s' \in S} p(s' \mid s_k, \mathcal{C}) ( c_{s_a} + V_{n-(k+1)}^{\pi}(s') )\label{eq:is1} \\
&=  \sum_{s' \in S} p(s' \mid s_k, \mathcal{C}) ((k+1)\cdot c_{s_a} + V_{n-(k+1)}^{\pi}(s') )\label{eq:is2}
\end{align}
On the other hand, the modified value of $s_a$ under policy $\pi'$ for $M'$ is given in~\eqref{eq:is3}, where $c_n(s_a, \pi'(s_a), s') = (k+1)\cdot c_{s_a} $ for any $s' \in S'$. 
\begin{align}
\mathbb{V}_n^{\pi}(s_a) & =   \sum_{s' \in S'} p'(s' \mid s_a, \pi'(s_a)) ( (k+1)\cdot c_{s_a} + \mathbb{V}_{n-(k+1)}^{\pi}(s') ) \label{eq:is3} 
\end{align}
Finally, by the induction hypothesis and the definition of the modified probability function $p'$, we reach that $V_n^{\pi}(s_a) $~\eqref{eq:is1} and $\mathbb{V}_n^{\pi}(s_a) $~\eqref{eq:is3}  are equivalent. Observing that $s_a$ was chosen arbitrarily concludes the proof.
\end{proof}

%% file: sections/corollary-proof.tex
\section{Proof of Corollary \ref{corol:value_eq}}

\begin{proof}Note that for any policy $\pi$ of $M$, it holds that $\pi(s) = \mathcal{C}$  when $s \in S - S'$.
For the sake of contradiction, assume that $\pi^\star$ from (33) 
is not optimal for $M$, i.e., there is a policy $\overline \pi$ such that $V^{\overline \pi}_n (s) < V^{\pi^\star}_n (s)$ for some $s \in S'$. Then by Proposition 4.1, the value of the projection of policy $\overline \pi$ on $S'$ is less than $\mathbb{V}_n^{\pi'^\star}$, which yields a contradiction. 
Likewise, the optimal value functions $V_n^{*}$ and $\mathbb{V}_n^{*}$ must also be equivalent.
\end{proof}